\documentclass[final, twocolumn, pra, superscriptaddress, floatfix, nofootinbib]{revtex4-2}
\usepackage[a4paper, margin=2cm]{geometry}
\usepackage[utf8]{inputenc}
\usepackage{xcolor}
\usepackage{graphicx}
\usepackage{physics}
\usepackage{float}
\usepackage{booktabs}
\usepackage{amssymb}
\usepackage[super]{nth}
\usepackage[caption=false]{subfig}
\PassOptionsToPackage{hyphens}{url}\usepackage[linktocpage=true, pdfencoding=auto, psdextra]{hyperref}
\usepackage{cleveref}
\usepackage{tikz}
\usepackage{qcircuit}
\usetikzlibrary{shapes,arrows,calc,positioning, automata,decorations.pathmorphing}
\hypersetup{
    colorlinks=false,       
    citecolor=blue,     
    filecolor=black,        
    linkcolor=red,      
    urlcolor=black,     
    linkbordercolor =red,   
    citebordercolor = blue, 
    urlbordercolor= blue    
}

\usepackage{xparse}
\NewDocumentCommand{\RZX}{ O{} O{} O{} }{R_{Z_{\scriptstyle #1} X_{\scriptstyle #2}}^\mathrm{\,#3}}
\NewDocumentCommand{\CNOT}{ O{} O{} }{\mathrm{CNOT}_{#1 #2}}
\NewDocumentCommand{\Pgate}{ O{} O{} }{P_{#1}^\mathrm{\,#2}}
\NewDocumentCommand{\Pdgate}{ O{} }{\dot P_{#1}}
\NewDocumentCommand{\Ptgate}{ O{} }{\tilde P_{#1}}
\NewDocumentCommand{\Phgate}{ O{} }{\hat P_{#1}}
\NewDocumentCommand{\Rdgate}{ O{} O{} }{\dot R_{Z_{#1} X_{#2}}}
\NewDocumentCommand{\AngleMap}{ O{} }{\tilde #1}

\usepackage{xspace}
\newcommand{\belem}{\texttt{ibmq\_belem}\xspace}
\newcommand{\brisbane}{\texttt{ibm\_brisbane}\xspace}

\begin{document}

\title{
Improving fidelity of multi-qubit gates using hardware-level pulse parallelization
}

\author{Sagar Silva Pratapsi}
\altaffiliation{Equal contribution.}
\author{Diogo Cruz}
\altaffiliation{Equal contribution.}
\affiliation{Instituto Superior Técnico, Universidade de Lisboa, Portugal}
\affiliation{Instituto de Telecomunicações, Portugal}


\begin{abstract}
Quantum computation holds the promise of solving computational problems which are believed to be classically intractable. However, in practice, quantum devices are still limited by their relatively short coherence times and imperfect circuit-hardware mapping. In this work, we present the parallelization of pre-calibrated pulses at the hardware level as an easy-to-implement strategy to optimize quantum gates. Focusing on $\RZX$ gates, we demonstrate that such parallelization leads to improved fidelity and gate time reduction, when compared to serial concatenation. As measured by Cycle Benchmarking, our most modest fidelity gain was from 98.16(7)\% to 99.15(3)\% for the application of two $\RZX(\pi/2)$ gates with one shared qubit. We show that this strategy can be applied to other gates like the CNOT and CZ, and it may benefit tasks such as Hamiltonian simulation problems, amplitude amplification, and error-correction codes.
\end{abstract}

\maketitle

\section{Introduction}\label{introduction}

The field of quantum computing has seen an explosion of interest in the recent decades~\cite{nature2022editorial, Bravyi2022, nielsenQuantumComputationQuantum2012}. Seminal results such as Shor's algorithm~\cite{shor_discrete_log, shor_factoring, nielsenQuantumComputationQuantum2012} and the threshold theorem for Quantum Error Correction~\cite{Devitt_2013,gottesman1997stabilizer,Ali07, aharonov1997fault} have provided the hope that quantum computers may one day solve problems that are currently intractable \cite{nielsenQuantumComputationQuantum2012}.

One major limitation to the field of quantum computation is that the quantum circuit abstraction breaks down when one uses real quantum devices. Physically, unitaries are only implemented approximately, using classical fields, and the resulting operation will not match the expected one with perfect fidelity \cite{glaserTrainingSchrodingerCat2015,kochChargeinsensitiveQubitDesign2007a}.
Furthermore, when designing quantum circuits, the circuit implementation is not tailored to some specific hardware. The circuit gates may be mapped sub-optimally to the pulse-level instruction schedule, thereby leading to longer pulse instructions and higher decoherence \cite{sundaresanReducingUnitarySpectator2020,kimHighfidelityThreequbitIToffoli2022,alexanderQiskitPulseProgramming2020,nguyenBlueprintHighPerformanceFluxonium2022,ibrahimEvaluationParameterizedQuantum2022}.

In fact, another main limitation of noisy intermediate-scale quantum (NISQ) devices is their relatively short coherence times~\cite{Preskill2018quantumcomputingin}. As the physical apparatus encoding the quantum state quickly interacts with the environment and suffers decoherence, there is a limited window where quantum operations can be performed on the quantum state, while still retaining acceptable fidelity. The rapid decoherence imposes limits on the circuit depths that are practically viable, and the number of gates applied, which limits the complexity of the quantum circuits that can be implemented in practice. This constraint ultimately limits the practical usefulness of quantum computing to solve problems where there is an expected theoretical quantum advantage. Consequently, in order to make the most use of current NISQ devices, we are compelled to shorten circuit times whenever possible, thereby enabling us to implement more complex circuits in the same time window.

Various approaches have been proposed to map abstract quantum circuits to efficient and fast pulse-level instructions. Transpilation, for instance, relies on gate cancellation and simplification, using optimization techniques \cite{namAutomatedOptimizationLarge2018}, and takes into account known connectivity constraints in the quantum device of interest, leading to a mapping that requires fewer intermediate gates to provide the desired connectivity, such as SWAP gates. At a lower level, circuit compilation \cite{earnestPulseefficientCircuitTranspilation2021,stengerSimulatingDynamicsBraiding2021} tries to decompose the original quantum circuit in terms of quantum gates that are less common but more representative of the underlying physical system. Finally, at the lowest level, we may use quantum control \cite{dalessandroIntroductionQuantumControl2021} and work with the pulse-level implementation directly, which might be the optimal policy to ensure high fidelities. However, this comes with significant trade-offs. The pulses that can be directly implemented in the superconducting qubits, for instance, do not map to commonly used quantum gates \cite{kochChargeinsensitiveQubitDesign2007a}. As a result, building large quantum circuits by using these pulse gates directly would be unintuitive and impractical.

In this work, we present an optimization strategy that lies between compilation and quantum control and that is readily available in contemporary devices---parallelization of pre-calibrated pulses. Following this procedure, we are able to map unitaries to native instructions, reduce gate times and increase fidelity, all with minimal effort. We show in \cref{sec:overview} that many Hamiltonian simulation problems can benefit from such a parallelization. One particular realization of this strategy relies on parallelizing $\RZX(\theta)$ gates with a shared qubit. In \cref{sec:implementation} we present the parallelization strategy and demonstrate that the Parallel $\RZX$ gate, $\Pgate[abc](\theta)$, is natively available in real devices and that it outperforms its serial counterpart. In fact, as measured by Cycle Benchmarking, by parallelizing two $\RZX(\pi/2)$ gates the fidelity goes from $98.16(7)\%$ to $99.15(3)\%$. We emphasize that the $\Pgate[abc](\theta)$ gates constitute a family of native high-fidelity three-qubit gates. Finally, in \cref{sec:applications}, we discuss how the parallelization strategy can be further generalized and applied to other problems.

\section{Pulse parallelization in NISQ Devices}\label{sec:overview}

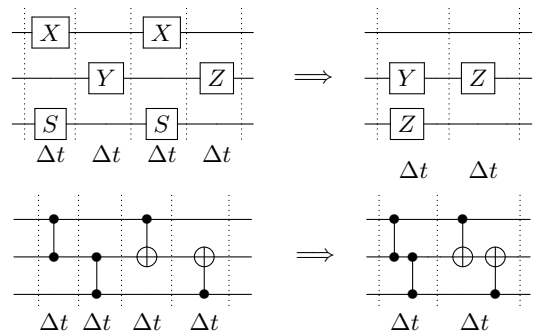
\begin{figure}
$$
    \Qcircuit @C=0.8em @R=0.6em {
        \ar@{.}[]+<0.5em,1em>;[d]+<0.5em,-2.5em>& \gate{X}\ar@{.}[]+<1em,1em>;[d]+<1em,-2.5em> & \qw\ar@{.}[]+<1em,1em>;[d]+<1em,-2.5em> & \gate{X}\ar@{.}[]+<1em,1em>;[d]+<1em,-2.5em> & \qw\ar@{.}[]+<1em,1em>;[d]+<1em,-2.5em> & \qw &\\
        & \qw & \gate{Y} & \qw & \gate{Z} & \qw & \push{\;\;\implies}\\
        & \gate{S} & \qw & \gate{S} & \qw & \qw& \\
        &\hspace{0ex}\Delta t&\hspace{0ex}\Delta t &\hspace{0ex}\Delta t&\hspace{0ex}\Delta t& &
        }\quad
    \Qcircuit @C=0.5em @R=0.6em @!R {
        & \qw\ar@{.}[]+<0em,1em>;[d]+<0em,-2.5em>& \qw & \qw\ar@{.}[]+<0.5em,1em>;[d]+<0.5em,-2.5em> & \qw & \qw & \qw & \qw\ar@{.}[]+<0em,1em>;[d]+<0em,-2.5em> & \qw\\
        & \qw & \gate{Y} & \qw & \qw & \gate{Z} & \qw & \qw & \qw\\
        & \qw & \gate{Z} & \qw & \qw & \qw & \qw & \qw & \qw\\
        &\hspace{7ex}\Delta t& &\hspace{9ex}\Delta t& & & & &
        }
$$
$$
    \Qcircuit @C=1.5em @R=1em {
        \ar@{.}[]+<1em,1em>;[d]+<1em,-2em>& \ctrl{1}\ar@{.}[]+<1em,1em>;[d]+<1em,-2em> & \qw\ar@{.}[]+<1em,1em>;[d]+<1em,-2em> & \ctrl{1}\ar@{.}[]+<1em,1em>;[d]+<1em,-2em> & \qw\ar@{.}[]+<1em,1em>;[d]+<1em,-2em> & \qw &\\
        & \control\qw & \ctrl{1} & \targ & \targ & \qw & \push{\implies}\\
        & \qw & \control\qw & \qw & \ctrl{-1} & \qw& \\
        &\hspace{0ex}\Delta t&\hspace{0ex}\Delta t &\hspace{0ex}\Delta t&\hspace{0ex}\Delta t& &
        }\quad
    \Qcircuit @C=0.5em @R=1em {
        & \qw\ar@{.}[]+<0em,1em>;[d]+<0em,-2em>& \ctrl{1} & \qw\ar@{.}[]+<1em,1em>;[d]+<1em,-2em> & \qw & \qw & \ctrl{1} & \qw\ar@{.}[]+<1em,1em>;[d]+<1em,-2em> & \qw & \qw\\
        & \qw & \control\qw & \ctrl{1} & \qw & \qw & \targ & \targ & \qw & \qw\\
        & \qw & \qw & \control\qw & \qw & \qw & \qw & \ctrl{-1} & \qw & \qw\\
        &\hspace{5ex}\Delta t& &\hspace{12ex}\Delta t& & & & & &
        }
$$
\caption{
\textbf{(Top)} Standard optimization techniques already simplify single qubit gate implementations, by canceling gates (as in the top qubit, where $X^2$ is simplified to the identity), merging them ($S^2 \rightarrow Z$), or simply parallelizing them at the pulse level (the pulse-level operations in the middle and bottom qubits go from being implemented in series to being in parallel).
\textbf{(Bottom)} While canceling and merging are also common techniques to simplify multi-qubit gates, parallelization is generally only performed if the gates share no qubits. In this work, we show that, by parallelizing the low-level pulses that define the complex multi-qubit gates, some overlapping gates can be implemented in parallel. As multi-qubit gates are the main contributors to the circuit duration, considerable time savings are possible, without significant overheads.}
\label{fig:diagram}
\end{figure}

Current quantum devices already heavily rely on parallelization of single qubit gates in order to achieve short circuit depths and durations (see \cref{fig:diagram}).
In this case, the pulse-level parallelization is straightforward, as there is no pulse overlap between qubits.
Similarly, multi-qubit gates can be parallelized at the pulse level if no qubits are shared among them.
However, in order to harness the full advantage of quantum devices, highly entangled quantum states are required, and to create these it is necessary to apply multi-qubit gates to shared qubits, for which parallelization at the pulse level is no longer straightforward.

Fortunately, while multi-qubit gates applied to overlapping qubits cannot be parallelized at the abstract circuit layer, we show that it is possible to parallelize them at the pulse layer, and that this parallelization is straightforward to implement once recognized.

Two gates $U_1, U_2$ may be parallelizable, for instance, if they are generated by two commuting time-independent Hamiltonians $H_1$ and $H_2$,
\begin{equation}
    U_1 U_2 
    =
    e^{-iH_1 t_1}
    e^{-iH_2 t_2}
    =
    e^{-it(H_1' + H_2')}.
    \label{eq:hamiltonian-parallelization}
\end{equation}
Here, $H_i' = H_i t_i / t$ are re-scaled versions of the Hamiltonians. The evolution on the right-hand side happens in time $t$, which may be chosen to be less than $t_1 + t_2,$ as illustrated in \cref{fig:diagram}. 

In actual quantum machines---using superconducting transmon qubits, trapped ions, or others---Hamiltonians are implemented via an external driving signal, like an electrical current or a laser \cite{kochChargeinsensitiveQubitDesign2007a}. Therefore, if two gates commute, it is possible that their underlying pulse signals can be superposed in order to approximate the evolution in \cref{eq:hamiltonian-parallelization}. Of course, it is important to emphasize that in real devices we can only approximate some desired Hamiltonian, as there are usually extra interactions that break commutativity \cite{sundaresanReducingUnitarySpectator2020,kimHighfidelityThreequbitIToffoli2022,alexanderQiskitPulseProgramming2020,nguyenBlueprintHighPerformanceFluxonium2022,ibrahimEvaluationParameterizedQuantum2022}. Additionally, implementing $H_1' + H_2'$ may consume more resources than $H_1$ and $H_2$ individually, and there will be practical limitations to its implementation, like voltage saturation or non-linear effects. Nevertheless, if $t$ is less than $t_1 + t_2$, parallelization may help achieve a computation in a shorter time interval, relative to the system's decoherence time, resulting in a net fidelity gain.

For this reason, it is interesting to ask whether pulse superposition can be used for gate parallelization in practice. We show that, in contemporary commercially-available quantum devices, the answer is positive.

\subsection{Hamiltonian simulation with \texorpdfstring{$\RZX$}{Rzx} gates} \label{sec:hamiltonian_simulation}

To motivate the use of parallel gates with an application, consider a general $n$-qubit Hamiltonian composed of Pauli terms of the form $h_i O_{i_1} \cdots O_{i_k}$ with $O \in \qty{X,Y,Z}$ and $\qty{i_1,\ldots,i_k}$ ($k\leq n$) an ordered subset of $\qty{1,\ldots, n}$, and $h_i$ real constants.
Since $\{I, X, Y, Z\}^{\otimes n}$ forms an operator basis, any Hamiltonian may in fact be decomposed in this way.
Using trotterization techniques \cite{ostmeyerOptimisedTrotterDecompositions2023}, we may implement the Hamiltonian evolution by evolving Pauli terms separately. 
For now, we analyze the simple case where two Pauli terms commute, which will allow us to parallelize their evolution following \cref{eq:hamiltonian-parallelization}.
Consider then, as an example, the three-qubit Hamiltonian acting on qubits $a, b, c,$
\begin{align}
    H &= h_1 O_a O_b + h_2 O_b O_c,\label{eq:H_2p2}\\
    \text{with } O &\in \qty{X,Y,Z}.
\end{align}
A simple change of basis at qubits $a,b,$ and $c$ may convert $H$ into
\begin{align}
    \tilde H &= U H U^\dagger\label{eq:U_single}\\
    &= h_1 Z_a X_b + h_2 X_b Z_c
\end{align}
where $U$ is the tensor product of single-qubit Pauli basis-change operators, given by the relation $\sqrt{X} Y  \sqrt{X}^\dagger = Z$ and its cyclical permutations under $(X, Y, Z)$.

An $R_{ZX}$ gate, acting on a control qubit $a$ and a target qubit $b$, is the unitary generated by $Z_a X_b$,
\begin{equation}
    \RZX[a][b](\theta) := \exp(-i (\theta /2) Z_a X_b).\label{eq:rzx-definition}
\end{equation}
By using $\tilde H$ and $U$, the evolution of $H$ in \cref{eq:H_2p2} can be implemented with the gate $\Pgate[abc](h_1t, h_2t)$, given by
\begin{align}
    \Pgate[abc](\theta_1, \theta_2)
    &: = \exp(-i\qty[
        \frac{\theta_1}{2}Z_aX_b
        + \frac{\theta_2}{2}X_bZ_c
    ]) \label{eq:przx} \\
    &= \RZX[a][b](\theta_1) \,
        \RZX[c][b](\theta_2).
    \label{eq:przx_2rzx}
\end{align}
For the sake of simplicity, we also define $\Pgate[abc](\theta) := \Pgate[abc](\theta, \theta)$. 
The $\RZX$ gate is then a candidate for parallelization.

As an application, consider the Heisenberg Hamiltonian
\begin{equation}
    H = X_1 X_2 + X_2 X_3 + Y_1 Y_2 + Y_2 Y_3 + Z_1 Z_2 + Z_2 Z_3.
\end{equation}
Applying the following Trotterization to its evolution
\begin{equation}
    \qty(
    e^{-i\frac{t}{n}(X_1X_2+X_2X_3)}
    e^{-i\frac{t}{n}(Y_1Y_2+Y_2Y_3)}
    e^{-i\frac{t}{n}(Z_1Z_2+Z_2Z_3)})^n
\end{equation}
reveals a case where the parallelization technique can be applied. Each term can be converted, using the single qubit gates from \cref{eq:U_single}, to the unitary
\begin{equation}
    e^{-i\frac{t}{n}(Z_1X_2+X_2Z_3)}, 
\end{equation}
which can be seen as a parallelized version of two $R_{ZX}(t/n)$ gates.

In the following section, we demonstrate that parallelizing two $\RZX$ gates indeed leads to a fidelity gain in current quantum devices. In fact, the $\RZX$ gate is quite versatile and can be used to construct other gates, such as the $\CNOT$ gate, which can also be parallelized, as we will explore in \Cref{sec:applications}.

Although we demonstrate our approach for the three-qubit case, it can be straightforwardly generalized to $(n+1)$ qubits. Considering one qubit to be the target $t$, and the other $n$ qubits the controls $c_i$, any Hamiltonian of the form
\begin{equation}
    H = \sum_{i=1}^n h_i O_{c_i} O_t\label{eq:H_ZX_n}
\end{equation}
can be directly implemented by a pulse-parallelized version of the gate
\begin{align}
    \Pgate[t,\vb c](\vb*{\theta}) &= \exp(-i\sum_{i=1}^n\frac{\theta_i}{2}Z_{c_i}X_t),\label{eq:PRZX_n}\\
    \text{with }\vb*{\theta} &= (\theta_1,\ldots,\theta_n) = (h_1t,\ldots,h_nt)\\
    \text{and }\vb c &= (c_1,\ldots, c_{n}).
\end{align}
Similarly, we define $\Pgate[t,\vb c](\theta) := \Pgate[t,\vb c](\theta,\ldots,\theta)$. 

So far, we have only considered the case where the target is the shared qubit. For generalized versions, see \cref{app:common_targets}.

We can now demonstrate that this parallelization strategy is feasible in contemporary quantum devices.

\section{Demonstration of a parallel \texorpdfstring{$\RZX$}{Rzx} gate} \label{sec:implementation}

In this Section, we demonstrate the implementation of $\Pgate[abc]$ (see \cref{eq:przx}) in a platform of superconducting qubits, IBM's \belem device.

We emphasize that the implementation we describe below relies on pre-calibrated pulses provided by the experimental platform. It thus requires minimal effort to set and is accessible on commercially available quantum devices.

\subsection{Transmon Hamiltonians}\label{sec:transmon}
The two-transmon system driven by the CR pulse at qubits $a$ and $b$ is approximately described \cite{alexanderQiskitPulseProgramming2020} by the time-independent Hamiltonian
\begin{align}
H_{\rm CR} &= \frac{Z_aB_b + I_a C_b}{2}\\
B_b &= \omega_{ZI} I_b + \omega_{ZX} X_b + \omega_{ZY} Y_b + \omega_{ZZ} Z_b\\
C_b &= \omega_{IX} X_b + \omega_{IY} Y_b + \omega_{IZ} Z_b,
\end{align}
where $I$ is the identity matrix and $X,Y,Z$ are the Pauli matrices. The symbols $\omega_{(\cdot)}$ are real coupling constants. We wish to isolate the main $Z_aX_b$ term. By carefully picking the phase of the CR pulse and applying a compensation pulse to the target qubit, undesirable terms in the Hamiltonian can be suppressed. Additionally, echo sequences can be employed to mitigate coherent errors (see \cref{app:echo_implementation}). 
A more detailed description of these noise suppression techniques can be found in \cite{sundaresanReducingUnitarySpectator2020}, and brief introductions in \cite{kimHighfidelityThreequbitIToffoli2022,alexanderQiskitPulseProgramming2020,nguyenBlueprintHighPerformanceFluxonium2022,ibrahimEvaluationParameterizedQuantum2022}. 
By suppressing these terms, we end up with the Hamiltonian
\begin{equation}
    H_{ZX} = \frac{Z_aX_b}{2},
\end{equation}
allowing us to implement the $\RZX[a][b](\theta)$ gate, given by \cref{eq:rzx-definition}.

Obtaining the pulse composition of $\RZX[a][b](\theta)$ for a general $\theta$ can be done from the calibrated pulses used for the CNOT gate, as explained in the next Section.

\begin{figure}[t]
    \centering
    \includegraphics[width=\linewidth]{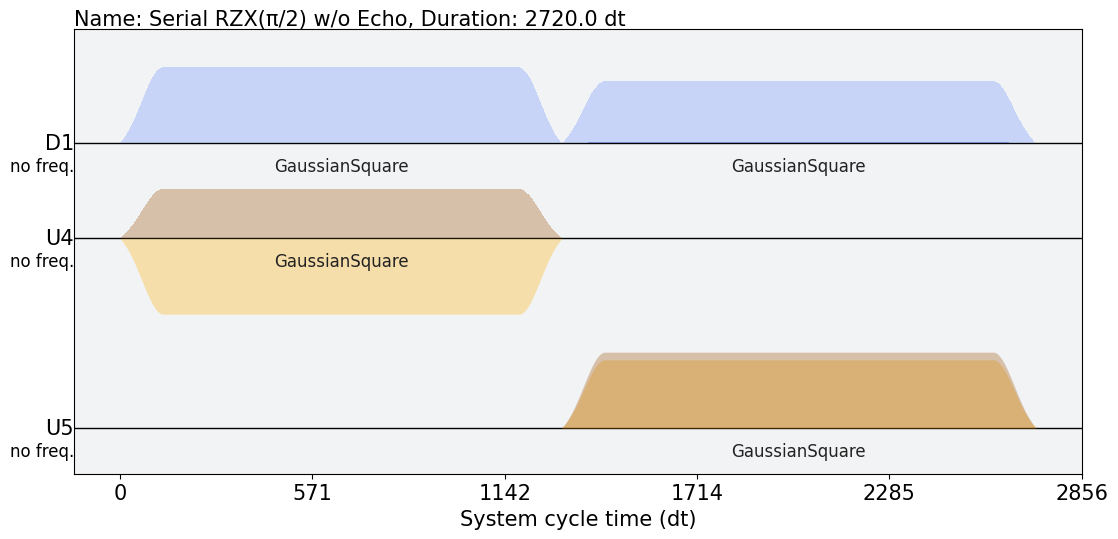}
    \includegraphics[width=\linewidth]{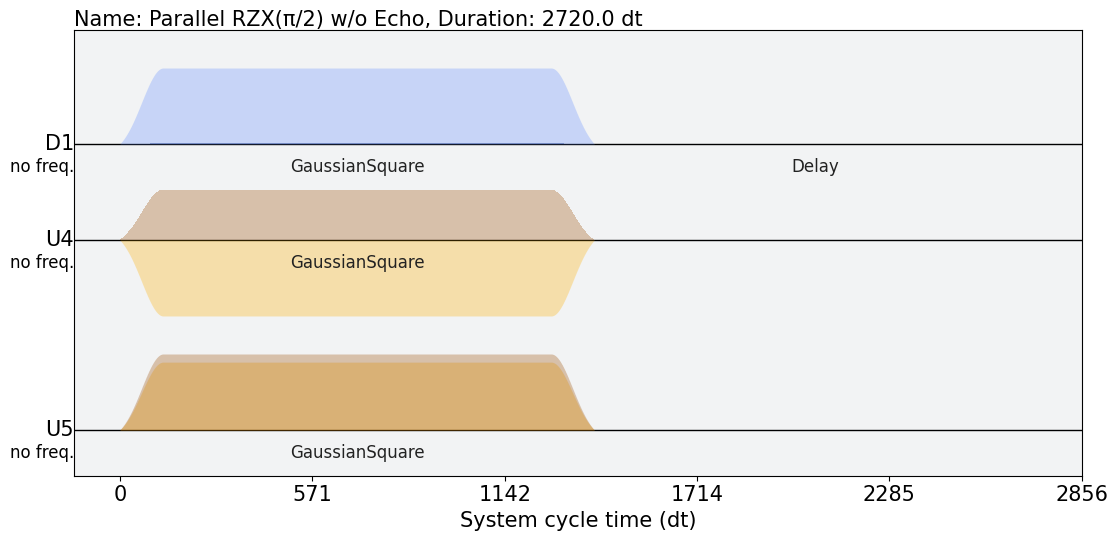}
    \caption{Example pulse sequence for the serial \textbf{(top)} and parallel \textbf{(bottom)} version of the double $\RZX$ gate, as described in \cref{sec:przx}, for the \belem device. By parallelizing the CR pulses (orange) and merging the compensation tones (blue), the gate duration of two $\RZX$ gates may be reduced by up to half. The parallelization can be made automatically, without additional calibration schemes, by using the $\RZX$ gate pulse shape stemming from the associated CNOT calibration, as described in \cref{sec:CNOT_calibration}.}
    \label{fig:pulse_schedule_noecho_belem}
\end{figure}

\subsection{Parallel \texorpdfstring{$\RZX$}{Rzx} pulse calibration and design}\label{sec:CNOT_calibration}

\subsubsection{Pulse calibration from existing \texorpdfstring{$\CNOT$}{CNOT} gates}\label{sec:RZX_calibration}
As the CNOT gate (with echo, see \cref{app:echo_implementation}) is widely used as a standard for 2-qubit entangling quantum gates, the calibration of current quantum devices generally attempts to mitigate the error present in CNOT gate applications. Since the echoed CNOT gates are implemented through the use of cross-resonance (CR) pulses with angle $\theta = \pm \pi/4$, pulses of this format are expected to present lower error rates, and can be used to define suitable pulses for other angle values.

Often, the echoed CNOT gate implementation is achieved by first decomposing the gates in terms of $R_{ZX}(\pi/4)$ (or $R_{ZX}(\pi/2)$ without echo) gates and additional single qubit gates. The error of the CNOT gates is then minimized by calibrating the pulses of the constituting gates, according to the decomposition used. There are several equivalent ways of decomposing CNOT gates in this manner. A possible decomposition for the unechoed CNOT gate is
\begin{align}
    \CNOT[c][t]
        & \quad\raisebox{0pt}{=}\quad X_c
        \RZX[c][t](\pi/2)
        S_c^\dagger
        Z_t
        \sqrt{X}_t
        Z_t,
    \label{eq:CNOT_decomposition_IBM} \\
\Qcircuit @C=0.5em @R=1.5em {
    & \ctrl{1} & \qw \\
    & \targ & \qw 
}
& \quad \raisebox{-8pt}{=} \quad
\Qcircuit @C=0.5em @R=0.5em {
    & \qw & \gate{S^\dagger} & \qw & \multigate{1}{\RZX(\frac{\pi}{2})} & \gate{X} & \qw \\
    & \gate{Z} & \gate{\sqrt{X}} & \gate{Z} & \ghost{\RZX(\frac{\pi}{2})} & \qw & \qw
} \nonumber
\end{align}
which is the one often used on IBM's quantum devices. An alternative formulation is
\begin{align}
    \CNOT[c][t]
        & \quad \raisebox{0pt}{=} \quad
        \RZX[c][t](\pi/2)
        S_c^\dagger
        X_t^{3/2}.
    \label{eq:CNOT_decomposition_simple}
    \\
\Qcircuit @C=0.5em @R=1.5em {
    & \ctrl{1} & \qw \\
    & \targ & \qw
}
& \quad \raisebox{-8pt}{=} \quad
\Qcircuit @C=0.5em @R=0.5em {
    & \gate{S^\dagger} & \multigate{1}{\RZX(\frac{\pi}{2})} & \qw \\
    & \gate{X^{3/2}} & \ghost{\RZX(\frac{\pi}{2})} & \qw
} \nonumber
\end{align}
The echoed variant (which may be directly used in the calibration) can be defined by using the echoed version of the $R_{ZX}(\pi/2)$ gate, as explained in \cref{app:echo_implementation}.

Finding calibrated pulses for the $\CNOT$ gate indirectly determines calibrated pulses for $\RZX(\pi/2)$, from which the pulses for $\RZX(\theta)$ (for $\theta$ other than $\pi/2$) can be set.

For $\RZX(\pi/2)$, in addition to the main cross-resonance pulse tones, there are also compensation tones applied to the target qubit during the main pulse. On IBM's devices, both types of pulses have the shape of a ``Gaussian square'', that is, a pulse that initially ramps up at $t_i$ following a squared exponential function until reaching a peak amplitude $A_{\rm CR}$, then plateaus at that amplitude for a set period of time, before ramping down to zero at $t_f$ using a similar squared exponential. See \cref{fig:layout,fig:pulse_schedule_noecho_belem} for examples, and \cite{IBMQuantumGaussianSquare,mckayQiskitBackendSpecifications2018} for a more thorough description. Let $t_{\rm CR} = |t_f-t_i|$ be the total pulse duration. The angle $\theta$ implemented by these pulses is proportional to the integral of the pulse. That is, if the pulse amplitude at time $t$ is $A(t)$, then
\begin{equation}
    \theta \propto \int_{t_i}^{t_f} A(t) \,\dd t.\label{eq:theta_from_int}
\end{equation}
If the ramping up and down portions are negligible, then this relation may be simplified to
\begin{equation}
    \theta \propto A_{\rm CR} t_{\rm CR}.
\end{equation}

Therefore, by modifying either the duration $t_{\rm CR}$ or the peak amplitude $A_{\rm CR}$ of the original $\theta=\pi/2$ pulse, we can modify the angle $\theta$ that is implemented as part of the $R_{ZX}$ gate. 
A more detailed description can be found at \cite{nguyenBlueprintHighPerformanceFluxonium2022,ibrahimEvaluationParameterizedQuantum2022}.

In this work, by default, we choose to keep the original peak amplitude and modify the pulse duration, in order to achieve the desired $\theta$ value, as it leads to lower gate durations.

\subsubsection{Pulse merging}\label{sec:przx}

We implement $\Pgate[abc](\theta_1, \theta_2)$ by merging the pulses that would constitute two separate $\RZX$ gates, instead of implementing each $\RZX$ gate in a serial manner, as in \cref{eq:przx_2rzx,fig:pulse_schedule_noecho_belem}. We thereby reduce the gate duration by up to half. The procedure can be implemented straightforwardly, if the pulse instructions for each $R_{ZX}(\theta)$ gate are previously known and well calibrated. These can be obtained following the standard procedure in the previous Section. The $\Pgate[abc](\theta_1, \theta_2)$ gate implementation is as follows:
\begin{enumerate}
    \item Schedule the two CR pulses to start running at the same time;
    \item (Optional) Stretch the gate duration of the shortest CR pulse to have the same duration as the longer pulse. Simultaneously, reduce the pulse amplitude $A_{\rm CR}$ so as to keep $\int_{t_i}^{t_f} A(t) \,\dd t$ constant. This step is not strictly required, but it was a firmware limitation present in pulse control in IBM's quantum devices at the time the results in this work were obtained;
    \item Merge the two compensation pulses. If the phase, peak amplitude, and duration of the original compensation pulses are $\phi_i, A_i, t_i$ (for the two pulses $i=1,2$), then the resulting combined compensation pulse has
    \begin{align}
        t &= \max\qty{t_1,t_2}\\
        A &= \frac{A_1 t_1 + A_2 t_2}{t}\\
        \phi &= \frac{\phi_1 A_1t_1 + \phi_2 A_2t_2}{A}.
    \end{align}
\end{enumerate}

\begin{figure*}
\centering
    \resizebox{8.4cm}{!}{\includegraphics{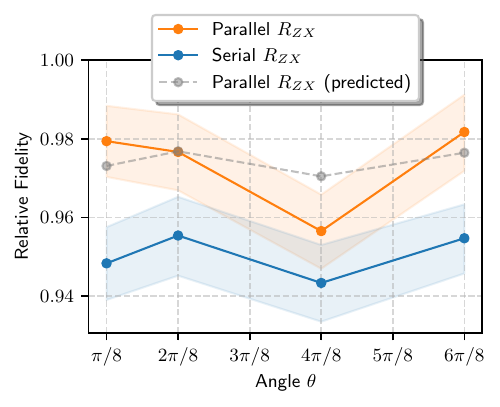}}
    \resizebox{8.4cm}{!}{\input{truth_table.tex}}
   \caption{
   \textbf{(Left)} Measured fidelities for the parallel and serial $\RZX$ gates, obtained via maximum-likelihood process tomography. The shaded regions correspond to the 90\% confidence intervals. To account for SPAM errors, the fidelities are presented relative to the fidelity of the identity process ($\theta = 0$), $F_\mathrm{id}=0.882\%$ (90\% confidence interval: [0.876, 0.888]). 
   In gray, we present the expected gain in fidelity that the Serial $\RZX$ would have by shortening its gate duration to that of the Parallel $\RZX$, using the decoherence model in \cref{eq:doherence-model}. We conclude that the Parallel $\RZX$ achieves a significantly better fidelity, compatible with coherence gains due to its shortened gate time.
   \textbf{(Right)}
  As an example, we show the parallel $\RZX(\pi/2)$ gate's logical table---probability of mapping a computational basis state $\ket{ijk}$ into $\ket{i'j'k'}$---reconstructed from its process tomography. To account for SPAM errors, the truth table was normalized as described in the main text. We observe results that are close to the ideal operation (black wire frame). See \cref{app:PTM} for a more complete characterization of the Parallel $R_{ZX}(\pi/2)$ using the Pauli Transfer Matrix. 
   }
    \label{fig:tomography}
\end{figure*}

Directly implementing $\Pgate[abc](\theta)$ using this procedure gives us a pulse schedule similar to that in \cref{fig:pulse_schedule_noecho_belem}. By doing so, it is as if we are implementing the two $R_{ZX}$ gates in succession.

This is not the only way to merge the two $\RZX$ gates. Here, we chose a gate duration corresponding to the duration of the longest CR pulse (that is, $t_{\rm CR} = \max\qty{t_1,t_2}$), but this duration can be chosen directly. For example, it is possible to choose a duration $\alpha t_{\rm CR}$ ($\alpha \in \mathbb R^+$) for the CR and compensation pulses, as long as the peak amplitudes are multiplied by a factor of $1/\alpha$. As a result, the quantity $\int_{t_i}^{t_f} A(t) \,\dd t$ is unchanged, so the same $\theta$ angle is implemented. Nonetheless, since shorter durations require higher amplitudes, for sufficiently high amplitude values the physical behavior of the system is non-linear, and leads \cref{eq:theta_from_int} to no longer be valid \cite{alexanderQiskitPulseProgramming2020}, and the implemented pulses will no longer correspond to the desired gate.
In practice, there is then a trade-off between using low pulse amplitudes and low gate durations. An analysis of such a trade-off is outside the scope of this work, since here we focus on reusing the pre-calibrated gates provided by the quantum devices.

The generalization of this procedure for $n$ $\RZX$ gates can be found in \Cref{app:n_RZX}.

\subsection{Gate benchmarking}\label{sec:gate_benchmarking}
To demonstrate the usefulness of the procedure we described in the previous Section, we implement a parallel $\RZX$ gate, $\Pgate[abc](\theta)$, on two pairs of qubits $(a, b)$ and $(c, b).$
We choose qubits $(a, b, c) = (2, 1, 3)$ of the \belem device (see \cref{app:hardware_configuration} for their characterization).

In practice, $\RZX$ gates on the \belem platform require echo pulses to suppress unwanted interactions. Furthermore, for $\abs{\theta} > \pi / 2$, $\Pgate[abc](\theta)$ can be reduced to $\Pgate[abc](\AngleMap{\theta})$ with $\lvert\AngleMap{\theta}\rvert < \pi/2.$
Our implementation of $\Pgate[abc]{}$ uses both of these techniques (see \cref{app:pulse-level}).

In \cref{fig:tomography} (left panel), we present the process fidelity for $\Pgate[abc]$ (``Parallel $\RZX$'') for several angles in the interval $[0, \pi].$
The fidelities were obtained from Maximum-Likelihood Estimation (MLE) process tomography~\cite{tomography_reference}.
To account for fluctuations in the calibration of \belem, we accumulated tomography data over three or more runs at different dates.
For comparison, we also present the fidelities for the serial variant $\RZX[a][b]\,\RZX[c][b]$ (``Serial $\RZX$'').
The shaded regions correspond to the 90\% confidence intervals following the procedure in Ref.~\cite{blume2012robust}.
To account for State Preparation and Measurement (SPAM) errors, the fidelities are presented relative to that of the identity operation, $F_\mathrm{id}=0.882$, which has a 90\% confidence interval of [0.876, 0.888].
Further results can be seen in \cref{app:cycle-benchmarking,app:PTM}.

We can see that the parallel $\RZX$ consistently achieves a better fidelity than its serial variant. In what follows, we argue that the better performance is compatible with the gain in coherence due to the shorter gate duration.

Current quantum devices are prone to decoherence \cite{nielsenQuantumComputationQuantum2012}, due to interactions that occur between the underlying quantum system and the macroscopic environment. As a result, when already accounting for other sources of noise, the fidelity over time is given by the exponential decay model \cite{hahnSpinEchoes1950}
\begin{equation}
    F(t) = (1-F_0)e^{-\beta t} + F_0,
    \label{eq:doherence-model}
\end{equation}
where $\beta$ is a device-dependent constant, which is lower for less noisy devices, and $F_0 = 2^{-n}$ is the fidelity of a maximally-mixed state of $n$ qubits. All else being equal, if the serial and parallel $\RZX$ implementations have a duration of $t_S$ and $t_P$, respectively, then their fidelities are given by $F_S = F(t_S)$ and $F_P=F(t_P)$, respectively, 
from which we obtain
\begin{equation}
    F_P = (1-F_0) \qty(\frac{F_S-F_0}{1-F_0})^{t_P/t_S} + F_0.
    \label{eq:decoherence-relative-scale}
\end{equation}
If $F_0$ is negligibly small (as is the case in the many-qubit setting), this expression can be simplified to
\begin{equation}
    F_P = F_S^{t_P/t_S}.
\end{equation}
In the best case scenario, where both $\RZX$ gates have the same duration, we have $t_S = 2t_P$, yielding
\begin{equation}
    F_P = \sqrt{F_S}.
\end{equation}
When considering the $(n+1)$-qubit formulation as in \cref{eq:PRZX_n}, we have $t_S = nt_P$, with which we obtain
\begin{equation}
    F_P = F_S^{1/n},
\end{equation}
yielding both a significant reduction in the circuit duration and a significant improvement in the gate fidelity.

In the left panel of \cref{fig:tomography}, we show in gray $F_S^{t_P/t_S}$, which appears to be close to the value of $F_\mathrm{P}$ within the experimental error. This supports the claim that the fidelity gains we observe are compatible with a gain in coherence.

In the right panel of \cref{fig:tomography} we show the reconstructed logical truth table for $\Pgate[abc](\pi / 2)$, that is, the probability $T_{ij}$ of obtaining a given computational basis $\ket j$ state when given as input another computational basis state $\ket i$, for $i,j=000,\ldots,111$.
To account for SPAM errors, the truth table is normalized so that its fidelity corresponds to the relative fidelity presented on the left panel, using the following procedure.
Let $\mathcal S_\mathrm{mle}$ be the Choi density operator of $\Pgate[abc](\pi/2)$ reconstructed via MLE tomography (with process fidelity $F_\mathrm{mle}$), and let $S_\mathrm{ideal}$ be the ideal Choi density operator of $\Pgate[abc](\pi/2)$.
We construct the normalized operator $S_\mathrm{norm}=\alpha S_\mathrm{mle} + (1-\alpha) S_\mathrm{ideal}$, for $\alpha \geq 0$, such that
\begin{align}
    F(S_\mathrm{norm}, S_\mathrm{ideal})
    & := \Tr{S_\mathrm{norm} \, S_\mathrm{ideal}} \notag\\
    & = \alpha F_\mathrm{mle} + (1-\alpha) \cdot 1 \notag\\
    & = F_\mathrm{mle} / F_\mathrm{id} \approx 0.957.
\end{align}
The bars in the figure correspond to 
\begin{equation}
    T^\mathrm{norm}_{ij} = \Tr{S_\mathrm{norm} \, (\rho_\mathrm{i}^T \otimes \rho_\mathrm{j})},
\end{equation}
where $\rho_i := \dyad i.$
The black wire frames correspond to the probabilities of the ideal process $\Pgate[abc](\pi/2).$

Finally, since process tomography does not discount SPAM errors, we characterize the SPAM-free performance of $\Pgate[abc]$ using Cycle Benchmarking. Usually, one uses Interleaved Randomized Benchmarking (IRB) \cite{PhysRevLett.106.180504, PhysRevLett.109.080505}, by interleaving applications of the desired operation with random Clifford gates.
However, IRB scales unfavorably. For this reason, we use Pauli operators instead of Clifford, in a process called \emph{Pauli twirling.} This results in a characterization procedure known as Cycle Benchmarking~\cite{erhardCharacterizingLargescaleQuantum2019}.

We use Cycle Benchmarking \cite{erhardCharacterizingLargescaleQuantum2019,wallmanNoiseTailoringScalable2016} to characterize the performance of the serial and parallel versions of $R_{ZX}(\pi/2)$, for all three-qubit Pauli channels. We use twirling depths $m\in \qty{4,8,16,32}$, and use 28 samples for each $m$. The fidelity $p_k$ of each Pauli channel $k$ is computed by fitting the fidelity decay using the model $Ap_k^m$ \cite{kimHighfidelityThreequbitIToffoli2022}. This enables us to isolate the gate fidelity, and disregard the state-preparation-and-measurement (SPAM) errors. As shown in \cref{fig:cycle_benchmarking} in \cref{app:cycle-benchmarking}, the average fidelity of the Parallel $R_{ZX}(\pi/2)$ thus obtained is $99.15(3)\%$ and that of the Serial $R_{ZX}$ is $98.16(7)\%,$ after normalizing by the fidelity of the identity process to discount for the intrinsic error of the Pauli twirling operators.

The difference in fidelities as measured with Cycle Benchmarking and MLE Tomography is explained by the facts that Cycle Benchmarking corrects for SPAM errors and does not compute a fidelity to some target gate. Cycle Benchmarking quantifies the error per interleaved Pauli, akin to IRB's error per Clifford.

\section{Applications}\label{sec:applications}

So far, we have showed how a pulse-parallelized version of $\RZX$ gates can be beneficial for the simulation of Hamiltonians given by \cref{eq:H_ZX_n}. In this Section, we show that, by extending the parallelization technique to CNOT and CZ gates, we can extend its use to simulating more general Hamiltonians, and implementing oracles.

\subsection{Parallelized CNOT gates} \label{sec:pcnot}

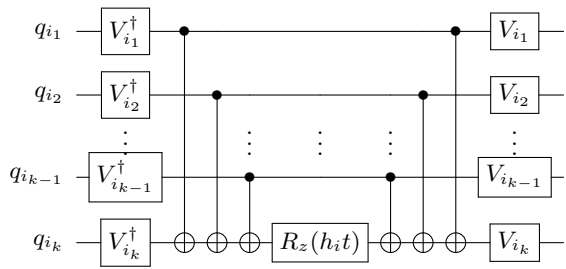
\begin{figure}[!htpb]
    \centering
    \scalebox{1.}{
        \Qcircuit @C=0.5em @R=0.7em {
        \lstick{q_{i_1}} & \gate{V_{i_1}^\dagger} & \ctrl{4} & \qw & \qw & \qw & \qw & \qw & \ctrl{4} & \gate{V_{i_1}} & \qw \\
        \lstick{q_{i_2}} & \gate{V_{i_2}^\dagger} & \qw & \ctrl{3} & \qw & \qw & \qw & \ctrl{3} & \qw & \gate{V_{i_2}} & \qw \\
        & \vdots & & & \vdots & \vdots & \vdots & & & \vdots & \\
        \lstick{q_{i_{k-1}}} & \gate{V_{i_{k-1}}^\dagger} & \qw & \qw & \ctrl{1} & \qw & \ctrl{1} & \qw & \qw & \gate{V_{i_{k-1}}} & \qw \\
        \lstick{q_{i_k}} & \gate{V_{i_k}^\dagger} & \targ & \targ & \targ & \gate{R_z(h_i t)} & \targ & \targ & \targ & \gate{V_{i_k}} & \qw \\
    }}
    \caption{Time evolution of the Pauli term $h_i O_{i_1} \cdots O_{i_k}$ using CNOT gates. We use qubit $i_k$ as the common CNOT target qubit, but any of the other $k-1$ qubits could be used.}
    \label{fig:pauli_term_evolution}
\end{figure}

\begin{figure}[t]
    \centering
$$
\Qcircuit @C=1.5em @R=0.5em {
    \lstick{q_{42}} & \ctrl{1}\ar@{.}[]+<1em,0.5em>;[d]+<1em,-6.5em>  & \qw\ar@{.}[]+<1em,0.5em>;[d]+<1em,-6.5em>       & \qw\ar@{.}[]+<1.55em,0.5em>;[d]+<1.55em,-6.5em>       & \qw\ar@{.}[]+<1.75em,0.5em>;[d]+<1.75em,-6.5em>           & \qw\ar@{.}[]+<1em,0.5em>;[d]+<1em,-6.5em>       & \qw\ar@{.}[]+<1em,0.5em>;[d]+<1em,-6.5em>       & \ctrl{1}  & \qw       \\
    \lstick{q_{43}} & \targ     & \ctrl{1}  & \qw       & \qw           & \qw       & \ctrl{1}  & \targ     & \qw       \\
    \lstick{q_{44}} & \qw       & \targ     & \ctrl{4}  & \qw           & \ctrl{4}  & \targ     & \qw       & \qw       \\
    \lstick{q_{47}} & \ctrl{1}  & \qw       & \qw       & \qw           & \qw       & \qw       & \ctrl{1}  & \qw       \\
    \lstick{q_{46}} & \targ     & \ctrl{2}  & \qw       & \qw           & \qw       & \ctrl{2}  & \targ     & \qw       \\
    \lstick{q_{54}} & \ctrl{1}  & \qw       & \qw       & \qw           & \qw       & \qw       & \ctrl{1}  & \qw       \\
    \lstick{q_{45}} & \targ     & \targ     & \targ     & \gate{R_z(t)} & \targ     & \targ     & \targ     & \qw       \\
    }
$$
$$
\Qcircuit @C=0.5em @R=0.5em {
    \lstick{q_{34}} & \ctrl{2}  & \qw\ar@{.}[]+<0.6em,0.5em>;[d]+<0.6em,-13em>       & \qw\ar@{.}[]+<0.6em,0.5em>;[d]+<0.6em,-13em>       & \qw       & \qw       & \qw\ar@{.}[]+<0.6em,0.5em>;[d]+<0.6em,-13em>       & \qw\ar@{.}[]+<1.7em,0.5em>;[d]+<1.7em,-13em>             & \qw       & \qw       & \qw\ar@{.}[]+<0.6em,0.5em>;[d]+<0.6em,-13em>       & \qw\ar@{.}[]+<0.6em,0.5em>;[d]+<0.6em,-13em>       & \qw       & \ctrl{2}  & \qw    \\
    \lstick{q_{42}} & \qw       & \ctrl{1}  & \qw       & \qw       & \qw       & \qw       & \qw             & \qw       & \qw       & \qw       & \qw       & \ctrl{1}  & \qw       & \qw    \\
    \lstick{q_{43}} & \targ     & \targ     & \ctrl{1}  & \qw       & \qw       & \qw       & \qw             & \qw       & \qw       & \qw       & \ctrl{1}  & \targ     & \targ     & \qw    \\
    \lstick{q_{44}} & \qw       & \qw       & \targ     & \ctrl{9}  & \qw       & \qw       & \qw             & \qw       & \qw       & \ctrl{9}  & \targ     & \qw       & \qw       & \qw    \\
    \lstick{q_{35}} & \ctrl{2}  & \qw       & \qw       & \qw       & \qw       & \qw       & \qw             & \qw       & \qw       & \qw       & \qw       & \qw       & \ctrl{2}  & \qw    \\
    \lstick{q_{48}} & \qw       & \ctrl{1}  & \qw       & \qw       & \qw       & \qw       & \qw             & \qw       & \qw       & \qw       & \qw       & \ctrl{1}  & \qw       & \qw    \\
    \lstick{q_{47}} & \targ     & \targ     & \ctrl{1}  & \qw       & \qw       & \qw       & \qw             & \qw       & \qw       & \qw       & \ctrl{1}  & \targ     & \targ     & \qw    \\
    \lstick{q_{46}} & \qw       & \qw       & \targ     & \qw       & \ctrl{5}  & \qw       & \qw             & \qw       & \ctrl{5}  & \qw       & \targ     & \qw       & \qw       & \qw    \\
    \lstick{q_{63}} & \ctrl{2}  & \qw       & \qw       & \qw       & \qw       & \qw       & \qw             & \qw       & \qw       & \qw       & \qw       & \qw       & \ctrl{2}  & \qw    \\
    \lstick{q_{65}} & \qw       & \ctrl{1}  & \qw       & \qw       & \qw       & \qw       & \qw             & \qw       & \qw       & \qw       & \qw       & \ctrl{1}  & \qw       & \qw    \\
    \lstick{q_{64}} & \targ     & \targ     & \ctrl{1}  & \qw       & \qw       & \qw       & \qw             & \qw       & \qw       & \qw       & \ctrl{1}  & \targ     & \targ     & \qw    \\
    \lstick{q_{54}} & \qw       & \qw       & \targ     & \qw       & \qw       & \ctrl{1}  & \qw             & \ctrl{1}  & \qw       & \qw       & \targ     & \qw       & \qw       & \qw    \\
    \lstick{q_{45}} & \qw       & \qw       & \qw       & \targ     & \targ     & \targ     & \gate{R_z(t)}   & \targ     & \targ     & \targ     & \qw       & \qw       & \qw       & \qw    \\
    }
$$
    \caption{Hamiltonian evolution of a Pauli term of the form $ZZ\ldots Z$. The qubit indices mark possible physical qubits in the \texttt{ibm\_brisbane} device that would enable these circuits to be directly implemented, without connectivity constraints. The dotted lines mark one time step $\Delta t$, for a total duration of $6\delta t$. We discount the $R_z$ gate, whose implementation is virtual and considered instantaneous. As an example, if the \texttt{ibm\_brisbane} device's decoherence is such that only circuits of duration $6\Delta t$ can be implemented with high fidelity, then \textbf{(Top)} using traditional techniques enables us to implement, at most, the 7-qubit term $\exp(-i t Z_1\ldots Z_7)$. \textbf{(Bottom)} However, using the parallelization techniques introduced in this work, for the same circuit duration, we can instead implement a 13-qubit term, thereby significantly extending the utility of low-depth circuits on current quantum devices.}
    \label{fig:CX_hamiltonian_simulation}
\end{figure}
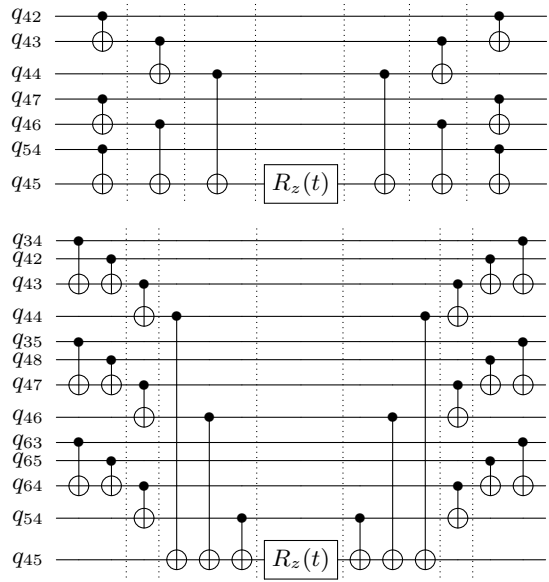

\begin{figure}[t]
    \centering
    \includegraphics[width=\columnwidth]{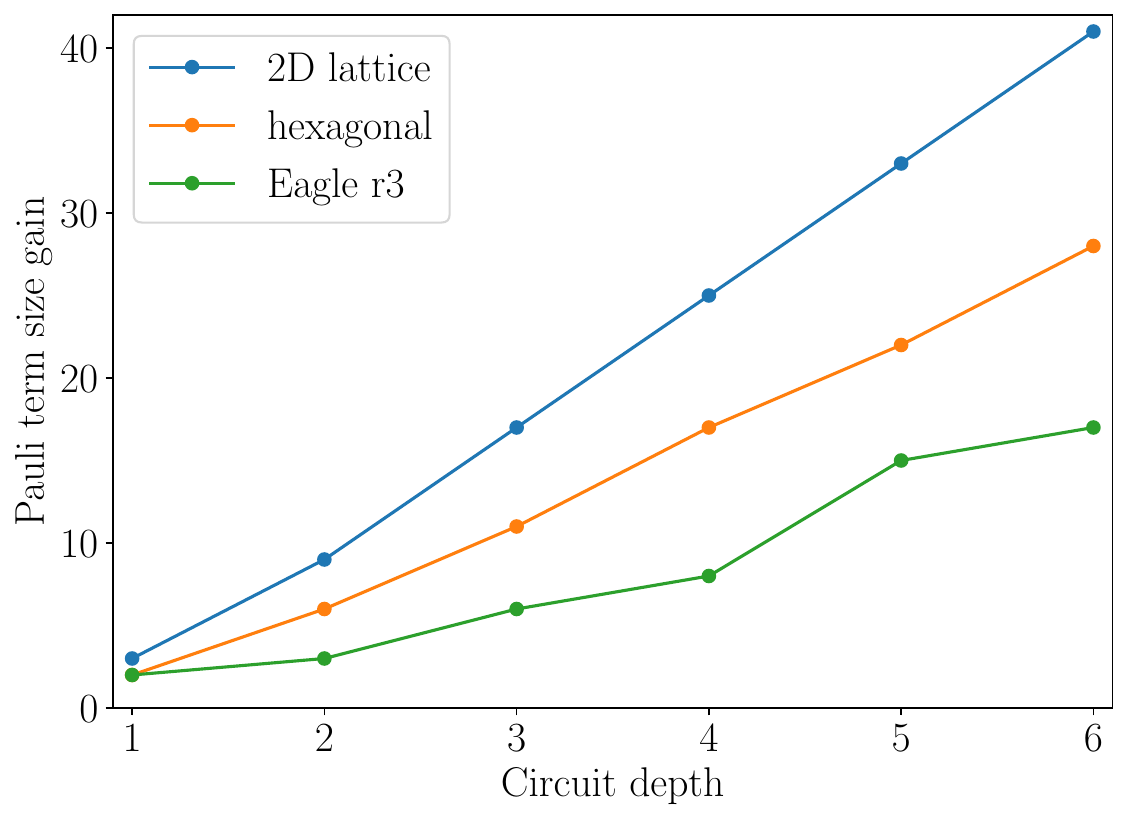}
    \includegraphics[width=0.49\columnwidth]{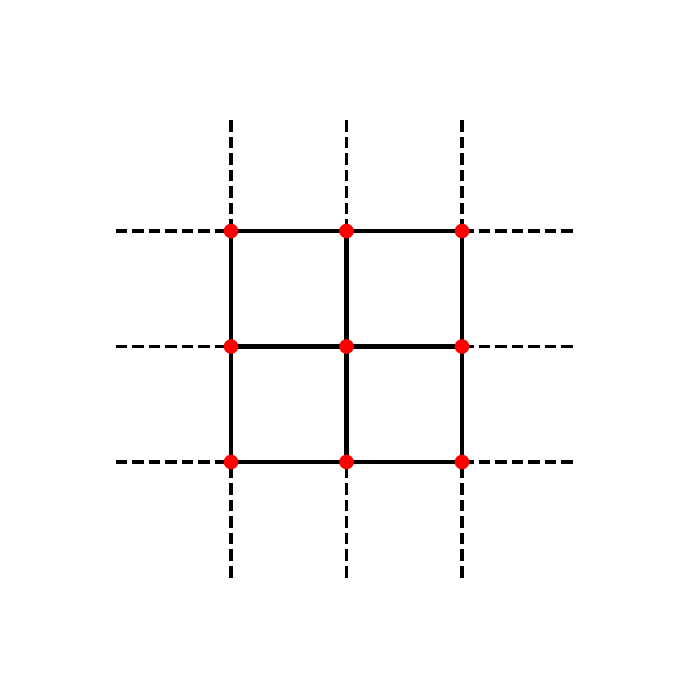}
    \includegraphics[width=0.49\columnwidth]{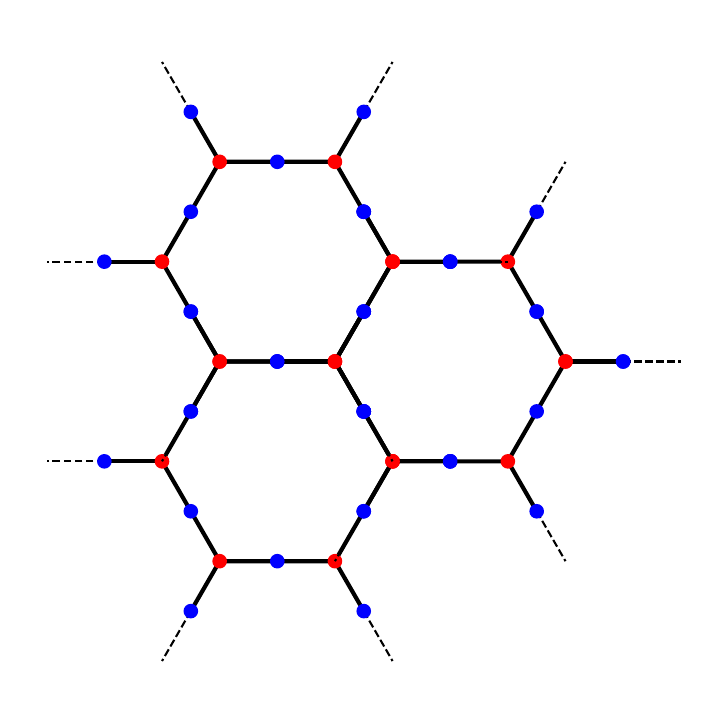}
    \caption{Parallelization benefit for different device layouts. \textbf{(Top)} Increase in the maximum qubit size of the simulatable Pauli-terms, when employing CNOT parallelization, assuming a fixed circuit depth. For example, in \cref{fig:CX_hamiltonian_simulation}, for depth 3, the simulatable Pauli term goes from being 7-qubit to 13-qubit in size, gaining 6 qubits. Among the layouts tested, for a 2D lattice configuration \textbf{(bottom left)}, the benefit is greatest, and it increases linearly with the circuit depth used. Configurations with weaker qubit connectivities, such as \textbf{(bottom right)} those of a standard hexagonal layouts (red dots stand for qubits) or those like IBM's Eagle r3 processor types (both red and blue dots stand for qubits) yield smaller, but still significant improvements. IBM's \texttt{ibm\_brisbane} possesses the latter layout.}
    \label{fig:device_layout}
\end{figure}

Using a similar approach to the $\Pgate[abc]$ implementation, and \cref{eq:CNOT_decomposition_simple}, we may apply several CNOT gates in parallel, at the pulse level. For $(n+1)$ qubits, we have
\begin{align}
    \prod_{c \in C} \CNOT[c][t]
    &= \Pgate[t,\vb c](\pi/2) (Q_t S_{c_1}^\dagger \cdots S_{c_n}^\dagger)\label{eq:CNOT_parallel}
\end{align}
where $Q = X^{3n/2 \text{ mod } 2}$, $C = \{c_1,\ldots, c_n\}$ is the set of $n$ control qubits, and there is a common target qubit $t$. For the generalized version with common control qubits, see \cref{app:common_targets}.

This type of parallel CNOT gate implementation proves particularly useful for Hamiltonian simulation. 
The evolution of an individual Pauli term $h_i O_{i_1} \cdots O_{i_k}$ (with $O \in \qty{X,Y,Z}$) can be implemented by first mapping it to the term $h_i Z_{i_1} \cdots Z_{i_k}$ through a basis change on each qubit $j$ (as in \cref{eq:U_single}) and then applying the evolution for that term (see \cref{fig:pauli_term_evolution}).

Following this procedure, the evolution of a Pauli term may be performed as in \cref{fig:pauli_term_evolution}, which relies on the application of many CNOT gates with the same target.

A possible strategy is then to evolve the $\CNOT$ gates in parallel. 
The main idea is that the $\CNOT$s may be written as a parallelization of $\RZX$ gates.
 
If the Pauli term of interest has $n+1$ components that are not the identity, then its evolution would require $n$ CNOT gates in succession before and after the $R_z$ gate. If the quantum device being used is such that all these CNOT gates share the same target at the physical level, then the duration of the CNOT gate implementation could be reduced by a factor of $1/n$, and the circuit fidelity significantly improved (see \cref{sec:gate_benchmarking}).

In practice, it is uncommon for current quantum devices, and especially those based on transmon qubits, to have the high qubit connectivities that would enable these significant improvements. For quantum devices based on superconducting qubits, each qubit can usually only be directly entangled with a small number of neighboring qubits. Entanglement to more distant qubits needs to be indirectly applied, through the use of SWAP gates. Current quantum devices tend to have qubits arranged on a lattice, with each at most directly connected to 3 neighboring qubits. More often, each qubit is only connected to 1 or 2 neighbors. As a result, for most situations, we can expect only 2 or 3 CNOT gates to be parallelized at the pulse level at a time.

Nonetheless, as can be seen in \cref{fig:CX_hamiltonian_simulation}, even with this practical limitation, CNOT gate parallelization can yield significant improvements. While the setup in \cref{fig:pauli_term_evolution} is the most straightforward way to implement a Pauli term evolution, there are other valid CNOT gate arrangements that result in valid circuits \cite{guiTermGroupingTravelling2021}. In general, if we wish to implement a Pauli term with $n$ non-identity components, we only need to ensure that the CNOT gates apply the gate $X^\alpha$ to the qubit with the $R_z$ gate, with $\alpha$ equal to the parity of the $n$ qubits with the $n$ non-identity components.

Without connectivity limitations, the optimal CNOT gate configuration permits CNOT gates applied with a circuit depth of $d$ to implement a Pauli term with $n=2^d$. However, as previously mentioned, current connectivity constraints lead the practical implementation to be far from optimal. For example, for the \texttt{ibm\_brisbane} device, a depth of 3 for the CNOT gate portion only allows the simulation of, at most, a 7-qubit Pauli term, and not 8, since the optimal implementation would require a qubit directly connected to 4 neighbors (see \cref{fig:CX_hamiltonian_simulation}). Nonetheless, using the parallelization techniques in our work, we increase the complexity of the simulatable Pauli terms, enabling the possibility of implementing a 13-qubit Pauli term using the same circuit depth. In essence, we gain access to 6 additional qubits for a fixed-depth simulation.

In general, the benefit gained from parallelizing CNOT gates will depend on the qubit connectivity layout of the qubit device. For a given depth $d$, layouts with more qubit-to-qubit connections will observe greater improvements in the number of additional qubits that can be included in the simulation, when parallelization is applied. See \cref{fig:device_layout} for an analysis of some common layouts.

\subsection{Parallelized CZ gates}

Similarly, we may implement CZ gates in parallel, by relying on the parallel CNOT setup in \cref{eq:CNOT_parallel}, yielding the implementation
\begin{align}
    \prod_{c \in C}\mathrm{CZ}_{ct}
    &= H_t \Pgate[t,\vb c](\pi/2) (Q_t H_t S_{c_1}^\dagger \cdots S_{c_n}^\dagger),\label{eq:CZ_parallel}
\end{align}
where $H$ stands for the Hadamard gate. Note that, since CZ is symmetric with respect to the 2 qubits it is applied to, here $t$ simply stands for the index of the shared qubit among the parallel CZ gates.

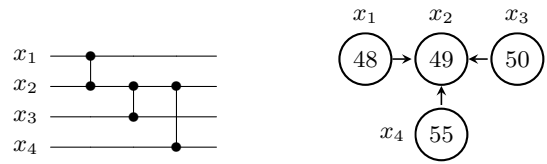
\begin{figure}[t]
\begin{minipage}{0.49\columnwidth}%
$$
    \Qcircuit @C=1.5em @R=1em {
        \lstick{x_1}& \ctrl{1} & \qw & \qw & \qw\\
        \lstick{x_2}& \control\qw & \control\qw & \control\qw & \qw\\
        \lstick{x_3}& \qw & \ctrl{-1} & \qw & \qw\\
        \lstick{x_4}& \qw & \qw & \ctrl{-2} & \qw
        }
$$
\end{minipage}%
\begin{minipage}{0.5\columnwidth}%
\begin{tikzpicture}[
        > = stealth, 
        shorten > = 1pt, 
        auto,
        node distance = 1cm, 
        semithick 
    ]

    \tikzstyle{every state}=[
        draw = black,
        thick,
        fill = white,
        minimum size = 2mm
    ]

    \tikzstyle{empty}=[
        draw = white,
        thick,
        fill = white,
        minimum size = 2mm
    ]

    \node[state] (0) {$48$};
    \node[state] (1) [right of=0] {$49$};
    \node[state] (2) [right of=1] {$50$};
    \node[state] (3) [below of=1] {$55$};
    \node[empty] (l0) [above=0cm of 0] {$x_1$};
    \node[empty] (l2) [above=0cm of 1] {$x_2$};
    \node[empty] (l1) [above=0cm of 2] {$x_3$};
    \node[empty] (l3) [left=0cm of 3] {$x_4$};

    \path[->] (0) edge node {} (1);
    \path[<-] (1) edge node {} (2);
    \path[<-] (1) edge node {} (3);
\end{tikzpicture}
\end{minipage}%
\caption{
\textbf{(Left)} Quantum circuit that implements the phase oracle $U = I - 2\sum_{i\in \iota_f}\dyad{i}{i}$, where $i$ is a bitstring and $i\in \iota_f$ iff $f(i)=1$, with $f(i)=f(x_1x_2x_3x_4)=x_1x_2+x_2x_3+x_2x_4$ mod 2. Different functions $f$ would lead to different placements for the CZ gates. In this case, since they share a common qubit ($x_2$), they can be parallelized at the pulse level. 
\textbf{(Right)} Possible implementation of the phase oracle in the \brisbane device, using physical qubits 48, 49, 50, and 55 to encode the variables $x_1$ to $x_4$, respectively. The arrows indicate the direction under which a direct $\RZX$ implementation is possible, with the arrow start (resp. end) corresponding to the $Z$ (resp. $X$) operator. Other qubit connections in \brisbane not relevant for the oracle implementation are not presented.}
\label{fig:phase_oracle}
\end{figure}

Among other uses, parallelizing CZ gates could prove useful to speed up the implementation of amplitude amplification. For example, a phase oracle whose encoded function corresponds to a Boolean formula, that in disjunctive normal form is a sum of terms of 2 variables, can be entirely implemented with CZ gates. See an oracle example in \cref{fig:phase_oracle}.

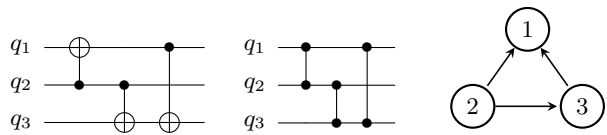
\begin{figure}[t]
\begin{minipage}{0.69\columnwidth}%
    $$\Qcircuit @C=1em @R=1em {
        \lstick{q_1}& \targ & \qw & \ctrl{2} & \qw\\
        \lstick{q_2}& \ctrl{-1} & \ctrl{1} & \qw & \qw\\
        \lstick{q_3}& \qw & \targ & \targ & \qw
        }\qquad\quad
    \Qcircuit @C=1em @R=1.3em {
        \lstick{q_1}& \ctrl{1} & \qw & \ctrl{2} & \qw\\
        \lstick{q_2}& \control\qw & \control\qw & \qw & \qw\\
        \lstick{q_3}& \qw & \ctrl{-1} & \control\qw & \qw
        }$$
\end{minipage}%
\begin{minipage}{0.3\columnwidth}%
\begin{tikzpicture}[
        > = stealth, 
        shorten > = 1pt, 
        auto,
        node distance = 1.1cm, 
        semithick 
    ]

    \tikzstyle{every state}=[
        draw = black,
        thick,
        fill = white,
        minimum size = 4mm
    ]

    \tikzstyle{noth}=[
        draw = white,
        thick,
        fill = white,
        minimum size = 4mm
    ]

    \tikzstyle{empty}=[
        draw = white,
        thick,
        fill = white,
        minimum size = 2mm
    ]

    \node[state] (1) {$1$};
    \node[state] (2) [below left=0.6cm and 0.3cm of 1] {$2$};
    \node[state] (3) [below right=0.6cm and 0.3cm of 1] {$3$};
    
    \path[<-] (1) edge node {} (2);
    \path[<-] (1) edge node {} (3);
    \path[->] (2) edge node {} (3);
\end{tikzpicture}
\end{minipage}%
\caption{\textbf{(Left)} With CNOT gates, the parallelization technique can only be applied to 2 of the CNOT gates if either the physical qubit $q_3$ (common target, \nth{2} and \nth{3} gates) or $q_2$ (common control, \nth{1} and \nth{2} gates, using technique in \cref{app:common_targets}) can be used as a common target for two $\RZX$ gates. Unfortunately, the parallelization technique is not applicable for some device layouts \textbf{(right)}, since then only configurations with a common target in $q_1$ are possible. \textbf{(Middle)} For CZ gates, the parallelization technique can be applied in theory with any qubit as the shared one and, for this particular example, to qubit $q_1$, thereby parallelizing the \nth{1} and \nth{3} gates.}
\label{fig:CZ_symmetry}
\end{figure}

The conclusions previously reached for the parallelization of CNOT gates mostly apply to CZ gates as well. The benefit of parallelizing CZ gates is greater the higher the qubit-to-qubit connectivity. However, unlike CNOT gates, CZ gates are symmetric, allowing more freedom to implement a desired circuit efficiently at the pulse level (see \cref{fig:CZ_symmetry} for an example).

\section{Discussion} \label{sec:discussion}

In this work, we presented a pulse parallelization technique that is readily available in NISQ devices.
Using IBM's \belem device, we demonstrated that the parallelization of $\RZX$ gates yields a fidelity advantage over their serial counterparts. As measured using Cycle Benchmarking, the minimum fidelity increase gained from parallelization was from 98.16(7)\% to 99.15(3)\%, for the Parallel $\RZX(\pi/2)$.

As the required pulses are straightforward to implement in most superconducting qubit architectures, including cloud-based-processors, the proposed approach has the potential to enable the simulation of larger circuits with higher fidelities, in NISQ devices. In particular, it seems promising for Hamiltonian simulation, as many quantum simulation problems can be readily decomposed in terms of $\RZX$ gates. Some well-known circuits, such as the encoding of error-correction codes~\cite{Devitt_2013}, can also benefit from the parallelization of many CNOT and CZ gates.

Different opportunities for further work present themselves. First, in our demonstration, we did not exploit the freedom to shorten the gate times. We focused on the simplest approach---to superpose the pulses as provided by the CNOT callibration---but it would be interesting to find out the minimal gate times before non-linear effects become relevant. Second, the parallelization could be tried in different architectures, like ion traps or cold atoms, which have different native Hamiltonians. Third, while we focused on $\RZX$ gates with shared target due to echo limitations, it would be interesting to see whether sharing the control qubit is feasible, or if one could implement parallel CZ gates. Finally, one could look into experimentally assessing the fidelity of many-qubit parallelization, starting with 3 parallel gates (see \cref{fig:device_layout,fig:phase_oracle} for examples available to current quantum devices).

Going beyond these generalizations, one could investigate the generation of arbitrary 3-qubit gates via either variational processes as outlined in \cite{kimHighfidelityThreequbitIToffoli2022} or via KAK decomposition \cite{earnestPulseefficientCircuitTranspilation2021,tucciIntroductionCartanKAK2005,vatanRealizationGeneralThreeQubit2004a,shendeSynthesisQuantumLogic2006,crossValidatingQuantumComputers2019,childsLowerBoundsComplexity2003}, and analyze if the additional freedom attained by controlling the pulses directly is useful.

In conclusion, our pulse parallelization technique demonstrates a path to higher fidelities and faster gate times on NISQ devices, with potential applications to quantum simulation, error correction, and broader quantum computation. Further work could focus on optimizing and generalizing this technique across different qubit architectures and larger qubit counts.

\begin{acknowledgments}

We would like to thank Duarte Magano and Miguel Murça for fruitful discussions.

SSP acknowledges the support from the ``la Caixa'' foundation, namely through scholarship No. LCF/BQ/DR20/11790030. SSP and DC acknowledges the support from FCT – Funda\c{c}\~{a}o para a Ci\^{e}ncia e a Tecnologia (Portugal) through scholarships 2023.01162.BD and UI/BD/152301/2021, respectively.

\end{acknowledgments}

\bibliography{main}

\appendix

\label{app:hardware_configuration}

\begin{table}[!htpb]
\begin{tabular}{lccc}
\toprule
                           & \multicolumn{3}{c}{\textbf{Qubits}} \\ \cmidrule{2-4}
\textbf{Property}          & \textbf{2}& \textbf{1}&\textbf{3} \\
\midrule
$T_1$ ($\mu$s)             & 69.3   & 78.0   & 53.6    \\
$T_2$ ($\mu$s)             & 38.6   & 63.8   & 56.5    \\
frequency (GHz)            & 5.36   & 5.25   & 5.17    \\
anharmonicity (GHz)        & -0.33  & -0.32  & -0.34   \\
readout error (\%)         & 2.4    & 3.2    & 3.4     \\
$p(1|0)$ (\%)              & 3.76   & 5.50   & 5.78    \\
$p(0|1)$ (\%)              & 1.04   & 0.90   & 1.02    \\
readout length (ns)        & 5351.1 & 5351.1 & 5351.1  \\
$\sqrt{X}$ infidelity (\%) & 0.030  & 0.041  & 0.067   \\
$\sqrt{X}$ duration (ns)   & 35.6   & 35.6   & 35.6    \\
CNOT(2,1) infidelity (\%)  & \multicolumn{3}{c}{1.1}   \\
CNOT(2,1) duration (ns)    & \multicolumn{3}{c}{412.4} \\
CNOT(3,1) infidelity (\%)  & \multicolumn{3}{c}{1.4}   \\
CNOT(3,1) duration (ns)    & \multicolumn{3}{c}{398.2} \\
\bottomrule
\end{tabular}
\caption{Device configuration.}
\label{tab:configuration}
\end{table}

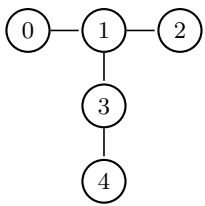
\begin{figure}[!htpb]
\centering
\begin{tikzpicture}[
        > = stealth, 
        shorten > = 1pt, 
        auto,
        node distance = 1cm, 
        semithick 
    ]

    \tikzstyle{every state}=[
        draw = black,
        thick,
        fill = white,
        minimum size = 4mm
    ]

    \node[state] (0) {$0$};
    \node[state] (1) [right of=0] {$1$};
    \node[state] (2) [right of=1] {$2$};
    \node[state] (3) [below of=1] {$3$};
    \node[state] (4) [below of=3] {$4$};

    \path[-] (0) edge node {} (1);
    \path[-] (1) edge node {} (2);
    \path[-] (1) edge node {} (3);
    \path[-] (3) edge node {} (4);
\end{tikzpicture}
\caption{Device layout. The nodes represent the qubits, and their labels, while the edges indicate the possible direct CNOT gate implementations.}
\label{fig:layout}
\end{figure}

\section{Device information}
The observed results were obtained on IBM's \belem device, in backend version 1.0.35. Qubits (2, 1, 3) were used, respectively, as (control, target, control), to obtain the tomography results in \cref{fig:cycle_benchmarking,fig:tomography,fig:pauli_transfer_matrix,fig:pauli_transfer_matrix_diff}. Its layout and a representative configuration are displayed in \cref{fig:layout,tab:configuration}, respectively. We note that, due to limitations in the number of circuits allowed to run concurrently in IBM's devices, and the large number of circuits required to obtain these results, IBM's device may have been recalibrated between some of the runs. Nonetheless, the device configuration did not deviate significantly from the values shown in \cref{tab:configuration}.

\section{Pulse-level optimizations}
\label{app:pulse-level}
In \cref{fig:pulse_schedule_noecho_belem}, we showed two possible pulse implementations for the concatenation of two $\RZX$ gates (top panel) and the parallel variant $\Pgate[abc]$ in \belem (bottom panel). For the purposes of this section, we will refer to these pulse implementations as $\Rdgate$ and $\Pdgate[abc],$ respectively.

Unfortunately, in practice, the error associated with $\Rdgate[a][b]$ and $\Pdgate[abc]$ is high, so we use two techniques to reduce their error: angle reduction and echo sequences.
This results in the gate implementation
\begin{align}
    \RZX[a][b](\theta)
    & := G_{ab}(\theta)
    \Rdgate[a][b]^\mathrm{\,echo}(\AngleMap{\theta})
    \nonumber
    \\
    \Pgate[abc](\theta)
    & := G_{b,(a,c)}(\theta)
    \Pdgate[abc]^\mathrm{\,echo}(\AngleMap{\theta}),
    \label{eq:pulse_sequences}
\end{align}
As explained in \cref{app:short_pulse}, the mapping $\theta \to \AngleMap{\theta}$ allows us to bring the angle of $\RZX$ to  the range $[-\pi/2, \pi/2]$. This mapping requires using the additional gates $G$.
On the other hand, $\Rdgate[][]^\mathrm{\,echo}$ and $\Pdgate[abc]^\mathrm{\,echo}$ refer to the echoed variants, defined in \cref{app:echo_implementation}.

In \cref{fig:echoed_gates}, we show the actual pulse sequences that we used for the serial and parallel versions of $\Pgate[abc](\pi / 2)$, using the decomposition in \cref{eq:pulse_sequences}.

\subsection{Angle reduction}
\label{app:short_pulse}
In the default implementation of $R_{ZX}(\theta)$ gates in IBM's devices, the duration of the CR pulse and compensation tones increases proportionally to the angle $\theta$ of the gate. As a result, for high values of $\theta$, the gate duration is substantial, leading to noticeable state decoherence. Moreover, the basis gates for the device are calibrated based on their performance for low $\theta$ (in particular, $\theta=\pm\pi/4$ for the CNOT gate with echo), resulting in an accumulation of coherent errors for the uncalibrated high $\theta$ values.
To circumvent this limitation, it is possible to construct the $\RZX$ (and $\Pgate[t,\vb c]$) gate for any $\theta$ value using only the pulse construction for $|\theta|\leq \pi/2$, without echo, or $|\theta|\leq \pi/4$, with echo. 

Since the $\RZX$ gate is periodic, so that $\RZX(\theta)=\RZX(\theta+2\pi)$, for the sake of simplicity, we consider that the input angle $\theta\in\mathbb R$ is pre-mapped to the range $(-\pi, \pi]$.

Let us map $\theta$ to the range $[-\pi/2, \pi/2]$ range using the function
\begin{align}
    \AngleMap{\theta} &:= \begin{cases}
    \theta-\text{sign}(\theta)\pi, & \text{if } \frac{\pi}{2}< |\theta|\leq \pi, \\
    \theta, & \text{if }|\theta|\leq \frac{\pi}{2}
    \end{cases}\\
    \AngleMap{\vb*{\theta}} &:= (\AngleMap{\theta}_1, \ldots, \AngleMap{\theta}_n).
\end{align}

We can then reduce $\RZX(\theta)$ to $\RZX(\AngleMap{\theta})$ by appending a correction gate $G$, because
\begin{equation}
    \RZX[a][b](\theta)
    = G_{ab}(\theta) \,
    \RZX[a][b](\AngleMap{\theta}),
\end{equation}
where
\begin{equation}
    G_{ab}(\theta) := \begin{cases}
        i(Z_a X_b), &\text{if }\frac{\pi}{2}< |\theta|\leq \pi\\
        I_aI_b,&\text{if }|\theta|\leq \frac{\pi}{2},
    \end{cases}
\end{equation}
so that we may only use angles $|\AngleMap{\theta}|\leq \frac{\pi}{2}$ in $\RZX$. The $i$ factor corresponds to a global phase and can be disregarded. Depending on the $\theta$ value, the additional gates $Z_a X_b$ may need to be applied. 

Similarly, we get can reduce the angle of $\Pgate[t,\vb c](\vb*{\theta})$, since it is a concatenation of $\RZX[c_i][t](\theta_i)$ gates,
\begin{align}
    \Pgate[t,\vb c](\vb*{\theta}) &=
        G_{t,\vb c}(\theta)
        \Pgate[t,\vb c](\AngleMap{\vb*{\theta}}),
\end{align}
where
\begin{align}
    G_{t,\vb c}(\theta)
        &:= G_{c_1 t}(\theta) \cdots G_{c_n t}(\theta) \label{eq:gate_G}\\
        &= Q_{\vb c}^Z(\vb*{\theta})\,
        Q_t^X(\vb*{\theta}).
\end{align}
In the last equality we separated the $Z$ and $X$ gates, since they simplify to
\begin{align}
    Q^Z_{\vb c}(\vb*{\theta}) &:= i^{\Delta(\vb*{\theta})} Z_{c_1}^{\delta(\theta_1)} \cdots Z_{c_n}^{\delta(\theta_n)}\\
    Q^X_{t}(\vb*{\theta}) &:= X_t^{\Delta(\vb*{\theta}) \text{ mod }2},
\end{align}
where we made use of the definitions
\begin{align}
    \delta(\theta) &:= \begin{cases}
        1, & \text{if }\frac{\pi}{2}< |\theta|\leq \pi\\
        0, & \text{if }|\theta|\leq \frac{\pi}{2}
    \end{cases}\\
    \Delta(\vb*{\theta}) &:= \sum_i \delta(\theta_i).
\end{align}
The $i^{\Delta(\vb*{\theta})}$ factor corresponds to a global phase and can be disregarded. 
When the $\theta$ angles are identical, the expression simplifies to
\begin{align}
    \Pgate[t,\vb c](\theta) &=
        Q_{\vb c}^Z(\theta) \,
        Q_t^X(\theta) \,
        \Pgate[t,\vb c](\AngleMap{\theta})
\end{align}
with
\begin{align}
    \Delta(\theta) &:= n\delta(\theta)\\
    Q^Z_{\vb c}(\theta) &:= (Z_{c_1} \cdots Z_{c_n})^{\delta(\theta)}\\
    Q^X_{t}(\theta) &:= X_t^{\Delta(\theta) \text{ mod }2}.
\end{align}
For even $n$, $Q^X_{t}(\theta)$ is always the identity, so no additional $X$ gates are applied.
Moreover, note that, in general, the implementation of the $Z$ gates is virtual, so it does not constitute an additional source of error.

\subsection{Echo sequences}
\label{app:echo_implementation}

Echo sequences are a basis-change technique to mitigate coherent errors \cite{kimHighfidelityThreequbitIToffoli2022,sundaresanReducingUnitarySpectator2020,alexanderQiskitPulseProgramming2020,nguyenBlueprintHighPerformanceFluxonium2022,ibrahimEvaluationParameterizedQuantum2022}.
We define echoed sequence for the $\RZX$ gate as
\begin{align}
    \RZX[a][b][echo](\theta)
    &:=
    X_a
    \RZX[a][b](-\textstyle\frac{\theta}{2})X_a
    \RZX[a][b](\textstyle\frac{\theta}{2}).
    \\ 
    \Qcircuit @C=0.5em @R=0.5em {
        & \multigate{1}{\RZX[][][echo]{(\theta)}} & \qw \\
        & \ghost{\RZX[][][echo]{(\theta)}} & \qw
    }
    & \;\; \raisebox{-8pt}{=} \;\;
    \Qcircuit @C=0.3em @R=0.3em {
        & \multigate{1}{\RZX(\frac{\theta}{2})} & \gate{X} & \multigate{1}{\RZX(-\frac{\theta}{2})} & \gate{X} & \qw \\
        & \ghost{\RZX(\frac{\theta}{2})} & \qw & \ghost{\RZX(-\frac{\theta}{2})} & \qw & \qw
    } \nonumber
\end{align}
For the serial double version of the $R_{ZX}$ gate, we apply the echo sequence to each gate separately,
\begin{align}
 &\RZX[c][b][echo](\theta)\RZX[a][b][echo](\theta)\\
    ={}& X_c\RZX[c][b](-\textstyle\frac{\theta}{2})X_c\RZX[c][b](\textstyle\frac{\theta}{2})\notag\\
    &\phantom{\qquad\qquad\qquad}X_a\RZX[a][b](-\textstyle\frac{\theta}{2})X_a\RZX[a][b](\textstyle\frac{\theta}{2}),
\end{align}
which is the gate we use in our implementation. By commuting the two middle $R_{ZX}$ gates, and some $X$ gates, we may also define an echoed version of the parallel variant. We obtain
\begin{multline}
 (X_aX_c)\RZX[c][b](-\textstyle\frac{\theta}{2})\RZX[a][b](-\textstyle\frac{\theta}{2})\\
 (X_aX_c)\RZX[c][b](\textstyle\frac{\theta}{2})\RZX[a][b](\textstyle\frac{\theta}{2}),
\end{multline}
yielding
\begin{align}
    \Pgate[abc][echo](\theta)
    :=
    (X_a X_c)
    \Pgate[abc](-\textstyle\frac{\theta}{2})
    (X_a X_c)
    \Pgate[abc](\textstyle\frac{\theta}{2})
\end{align}
\begin{align}
    \Qcircuit @C=0.5em @R=0.5em {
        & \multigate{2}{\Pgate[][echo](\theta)} & \qw \\
        & \ghost{\Pgate[][echo](\theta)} & \qw\\
        & \ghost{\Pgate[][echo](\theta)} & \qw
    }
    \;\; \raisebox{-14pt}{=} \;\;
    \Qcircuit @C=0.5em @R=0.3em {
        & \multigate{2}{\Pgate(\frac{\theta}{2})} & \gate{X} & \multigate{2}{\Pgate(-\frac{\theta}{2})} & \gate{X} & \qw \\
        & \ghost{\Pgate(\frac{\theta}{2})} & \qw & \ghost{\Pgate(-\frac{\theta}{2})} & \qw & \qw\\
        & \ghost{\Pgate(\frac{\theta}{2})} & \gate{X} & \ghost{\Pgate(-\frac{\theta}{2})} & \gate{X} & \qw
    } \nonumber
\end{align}

\begin{figure*}[!htpb]
    \centering
    \includegraphics[width=.48\linewidth]{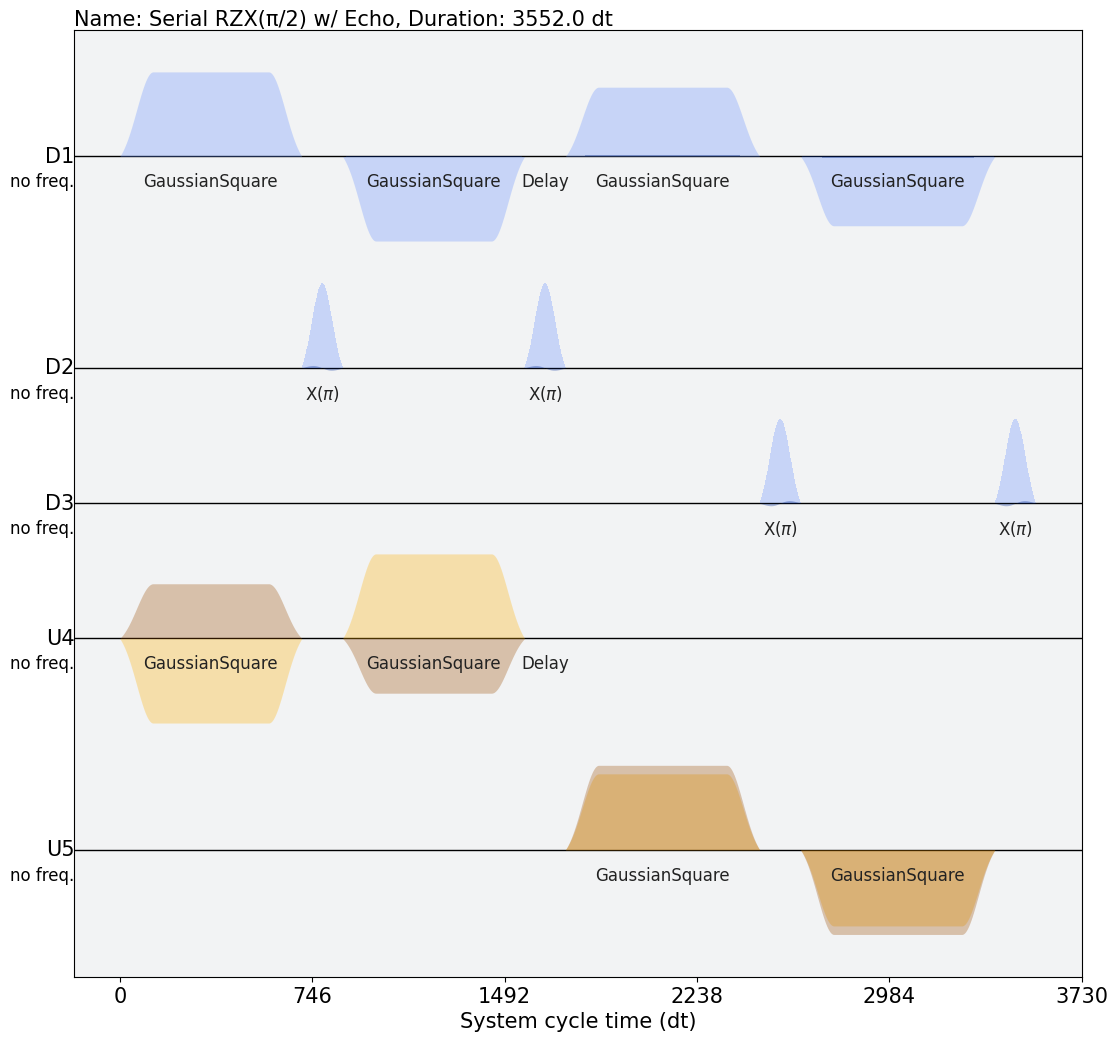}
    \includegraphics[width=.48\linewidth]{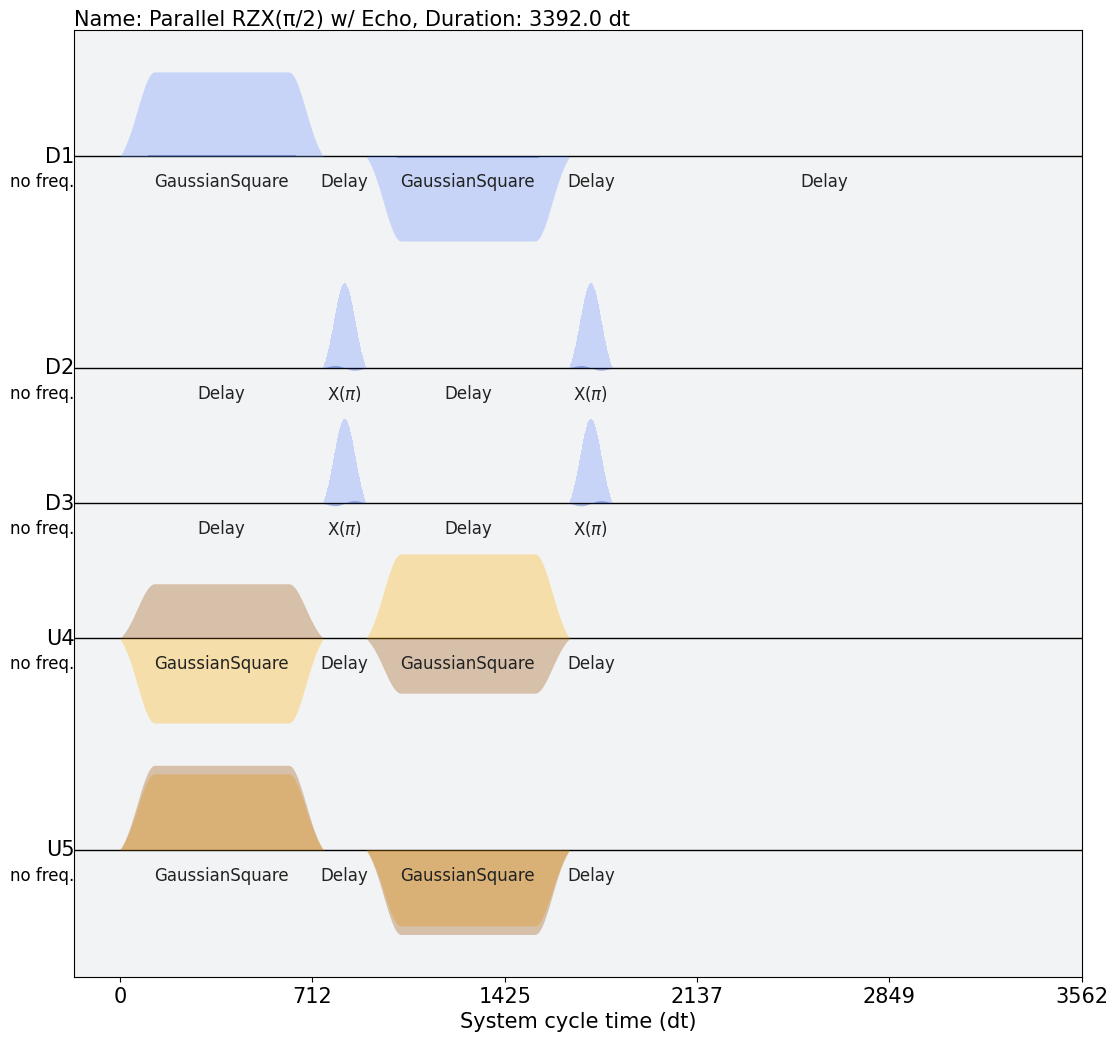}
    \caption{Example pulse sequence for the \textit{echoed} serial \textbf{(left)} and parallel \textbf{(right)} versions of the double $\RZX(\pi/2)$ gate, for the \belem device, as described in \cref{app:pulse-level}. As with the unechoed version (c.f. \ref{fig:pulse_schedule_noecho_belem}, the gate duration can be reduced by up to half by using the parallel version. For $\theta=\pi/2,$ the gate $G$ in \cref{eq:gate_G} is equal to the identity operation.}
    \label{fig:echoed_gates}
\end{figure*}

An example of the pulses associated with these echoed gates can be seen in \cref{fig:echoed_gates}.

The echoed implementation of the parallel gate generalizes to $n$ qubits and different $\theta$ angles, yielding
\begin{equation}
    \Pgate[t,\vb c][echo](\vb*{\theta})
    :=
    X_{\vb c}
    \Pgate[t,\vb c](-\textstyle\frac{\vb*{\theta}}{2})
    X_{\vb c}
    \Pgate[t,\vb c](\textstyle\frac{\vb*{\theta}}{2}),
\end{equation}
where $X_{\vb c}$ denotes the application of $X$ gates to all $(n-1)$ control qubits.

\section{Parallel \texorpdfstring{$\RZX$}{Rzx} gates with common control qubits}\label{app:common_targets}

In this work, we have focused on a parallelized version of the $\RZX$ gate where the target qubit is shared, yielding the gate in \cref{eq:PRZX_n}. Using a similar reasoning, it is possible to define the gate
\begin{align}
    \Pgate[c,\vb t](\vb*{\theta}) &:= \exp(-i\sum_{i=1}^n\frac{\theta_i}{2}Z_{c}X_{t_i}),\\
    \text{with }\vb t &= (t_1,\ldots, t_{n}),
\end{align}
where $n$ $\RZX$ gates use a shared control qubit and distinct target qubits. Under this approach, the CR pulse parallelization would proceed as previously described, but there would be no need for merging compensation pulses, since each of these pulses would be applied to a separate target qubit.

Unfortunately, this approach does not prove as fruitful as the one with the common target qubit, since the reasoning behind the application of echo, as explained in \cref{app:echo_implementation}, no longer holds. Consequently, suppressing some of the undesired Pauli terms in the CR Hamiltonian becomes non-trivial, and the naïve pulse parallelization achieves poor fidelity.

Nonetheless, a high-fidelity implementation can be done by using the common-target gate, as given by
\begin{align}
    \Ptgate[c,\vb t](\vb*{\theta}) = H_c H_{\vb t} \Pgate[c,\vb t](\vb*{\theta}) H_c H_{\vb t},
\end{align}
which can take advantage of the echo implementation of the $\Pgate[t,\vb c](\vb*{\theta})$ gate.

If we consider the most general case
\begin{align}
    \Phgate[\vb c,\vb t](\vb*{\theta}) &:= \exp(-i\sum_{i=1}^n\frac{\theta_i}{2}Z_{c_i}X_{t_i}),\\
    \text{with }\vb t &= (t_1,\ldots, t_{n}), \quad \vb c = (c_1,\ldots, c_{n}),
\end{align}
we run into similar issues as before, since there is not a straightforward way to implement echo for the terms that share control qubits. Moreover, we may also have qubits that act both as a control and a target, depending on the Pauli term. In general, such components are not possible to parallelize, as they can be used to build gates that cannot be implemented with a single CR pulse. 

In order to still take advantage of parallelization in this setting, a suboptimal but simple method is to partition the $\Phgate[\vb c,\vb t]$ gate into a product of $P$ and $\tilde P$ gates, and parallelize each of these gates separately. It may be possible to improve on this method by relying on more complex circuit optimization techniques, but these are outside the scope of this work.

Given the straightforward connection between $\RZX$ gates and CNOT and CZ gates, the same principles can be applied to parallelize CNOT and CZ gates, when the gates have some control qubits in common.

\section{Pulse parallelization for \texorpdfstring{$n$ $\RZX$}{n Rzx} gates}\label{app:n_RZX}

It is straightforward to generalize the procedure in \cref{sec:przx} in order to parallelize $n$ $\RZX$ gates:

\begin{enumerate}
    \item Schedule the $n$ CR pulses to start running at the same time.
    \item (Optional) Stretch the gate duration of the shorter CR pulses to have the same duration as the longest pulse. Simultaneously, reduce the pulse amplitude $A_{\rm CR}$ so as to keep $\int_{t_i}^{t_f} A(t) \,\dd t$ constant. 
    \item Merge the $n$ compensation pulses. If $n$ is large, the computed amplitude may be higher than the maximum amplitude $A_{\rm max}$ below which \cref{eq:theta_from_int} stills holds, so the gate duration may need to be increased to compensate. If the phase, peak amplitude, and duration of the original compensation pulses are $\phi_i, A_i, t_i$ (for the $n$ pulses $i\in\qty{1,\ldots, n}$), then the resulting combined compensation pulse has parameters
    \begin{align}
        S &:= \sum_i A_i t_i\\
        t &= \max\qty{S/A_{\rm max}, \max_i t_i}\\
        A &= \min\qty{A_{\rm max}, \frac{S}{t}}\\
        \phi &= \frac{1}{A}\sum_i \phi_i A_it_i.
    \end{align}
\end{enumerate}

\section{Characterization of parallel implementation}
To better characterize our implementation of the parallel gate $\Pgate[abc],$ in this section we present a more complete experimental characterization of $\Pgate[abc](\pi/2).$ We do this because $\Pgate[abc](\pi/2)$ is a Clifford operator, so we can apply Cycle Benchmarking to characterize its fidelity (see the following sub-section). As we can see from \cref{fig:tomography}, this is also the gate that presents the highest errors, so it serves as a lower bound for the other angles.

\subsection{Cycle Benchmarking}\label{app:cycle-benchmarking}
\begin{figure*}[!htpb]
    \centering\includegraphics{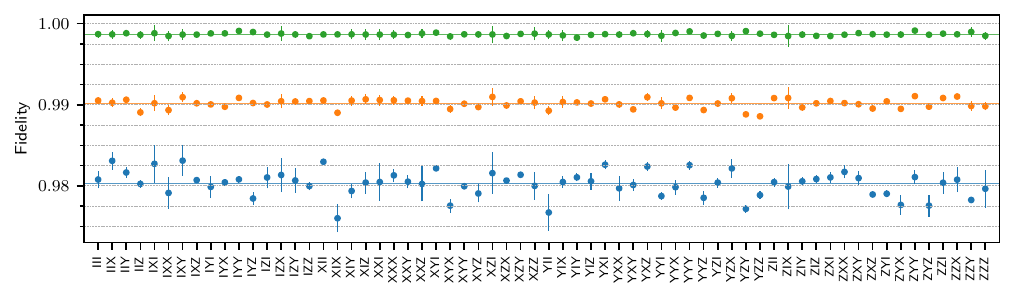}
    \caption{Fidelities obtained from Cycle Benchmarking for the identity (green) and the parallel (orange) and serial (blue) version of two $\RZX(\pi/2)$ gates, for all the different Pauli channels.
    The horizontal lines correspond to the respective average fidelities. 
    We obtain fidelities of $99.15(3)\%$ and $98.16(7)\%$, respectively, for the Parallel and Serial $\RZX$ gates. These fidelities were re-scaled by that of the identity operation, whose fidelity is $99.869(8)\%$, to discount the error in the Pauli twirling operations.
    As expected from the decoherence model~\cref{eq:decoherence-relative-scale}, we observe that $F_{\rm parallel} \simeq F_{\rm serial}^{0.514}=99.05(3)\%$, with $t_S = 0.40~\mathrm{\mu s}$, $t_P = 0.79~\mathrm{\mu s}$, and $0.514=t_P/t_S$ being the duration of the parallel gate relative to that of the serial gate. A similar result is obtained by using \cref{eq:decoherence-relative-scale}, since $F_0$ is small. Moreover, we also observe that the fidelity uncertainty is noticeably smaller in the parallel version, reflecting more predictable behavior from the parallel gate.}
    \label{fig:cycle_benchmarking}
\end{figure*}

As mentioned in the main text, we ran Cycle Benchmarking \cite{erhardCharacterizingLargescaleQuantum2019,wallmanNoiseTailoringScalable2016} for all three-qubit channels with depth $m\in \qty{4,8,16,32}$ and 28 samples for each $m$. The fidelity $p_k$ of each Pauli channel $k$ is extracted via a fit, assuming that the measured fidelity decays as $Ap_k^m$ \cite{kimHighfidelityThreequbitIToffoli2022}. This enables us to isolate the gate fidelity, and disregard the state-preparation-and-measurement (SPAM) errors. As the computed fidelities include the error contributions from Pauli-twirling, we also run the same circuits but without the tested gate, to compute the fidelity $p_k^{(\rm ref)}$ associated with the Pauli-twirling component alone. Consequently, the resulting gate fidelities after deducting the SPAM and Pauli-twirling errors are given by $p_k/p_k^{(\rm ref)}$. The obtained fidelities are shown in \cref{fig:cycle_benchmarking}.

\subsection{Pauli Transfer Matrix}\label{app:PTM}
We provide a visualization of our implementation of the parallel $R_{ZX}$ operation, $\Pgate[abc]$, for the angle $\theta = \pi/2,$ by reconstructing its Pauli Transfer Matrix (PTM). To do so, we begin with the reconstruction from Maximum-Likelihood Estimation (MLE) Process Tomography~\cite{tomography_reference}, and subsequently calculate the PTM.

Although quantum processes are in general operators between complex vector spaces, it is possible to fully describe them with a real matrix whose elements belong to the real interval $[-1, 1],$ using the so-called \emph{Pauli Transfer Matrix}~\cite{greenbaum2015introduction}.
The elements of the Pauli Transfer Matrix $T$, associated with the quantum process $\Lambda,$ are given by
\[
    T_{ij} = \Tr{P_j \, \Lambda(P_i)},
\]
where $P_i$ and $P_j$ are the Pauli strings taken from $\{I, X, Y, Z\}^{\otimes 3},$ ordered lexicographically. Here, $I$ is the $2 \times 2$ identity and $X,Y,Z$ are the Pauli operators. The PTM of our three-qubit process is therefore a $4^3 \times 4^3 = 64 \times 64$ real matrix.

From the Pauli Transfer Matrix, we can infer several properties of the operator $\Lambda$. The trace-preserving condition is met if and only if the first row is 1 followed by all zeroes, i.e., $T_{0j} = \delta_{0j}.$ If $\Lambda$ is unital (as unitary matrices are), then the first column must also be 1 followed by all zeroes. The PTMs $T_{\chi}$ of general quantum processes $\chi$ also multiply, that is, $T_{\Lambda \circ \chi} = T_{\Lambda} T_{\chi}$. For more information on the PTM see, for example, Ref.~\cite{greenbaum2015introduction}.

\begin{figure*}[!htpb]
    \centering
    \includegraphics[width=1\textwidth]{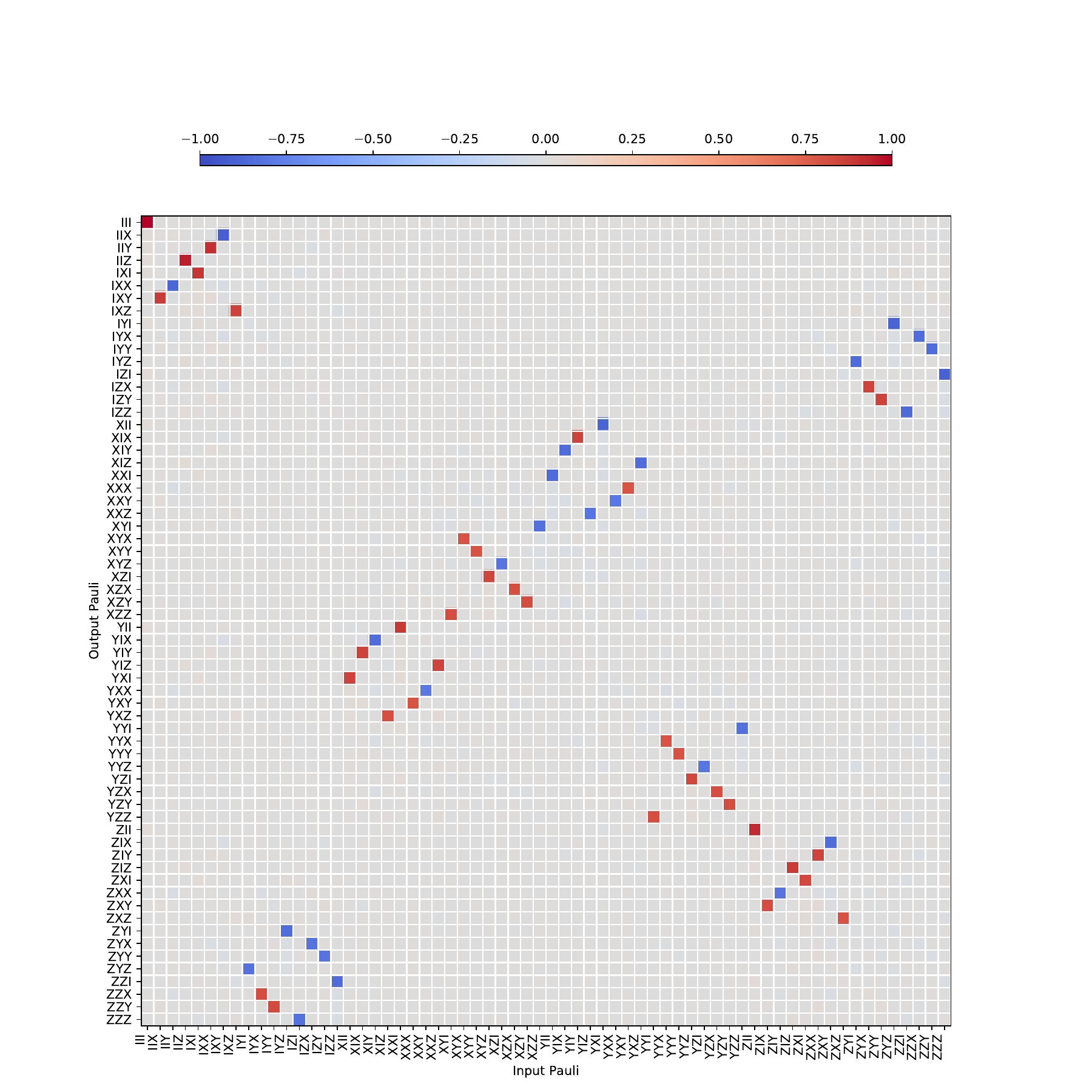}
    \caption{Pauli Transfer Matrix of the Parallel $R_{ZX}(\pi / 2)$, reconstructed from Process Tomography, as discussed in the main text.}
    \label{fig:pauli_transfer_matrix}
\end{figure*}

\begin{figure*}[!htpb]
    \centering
    \includegraphics[width=1\textwidth]{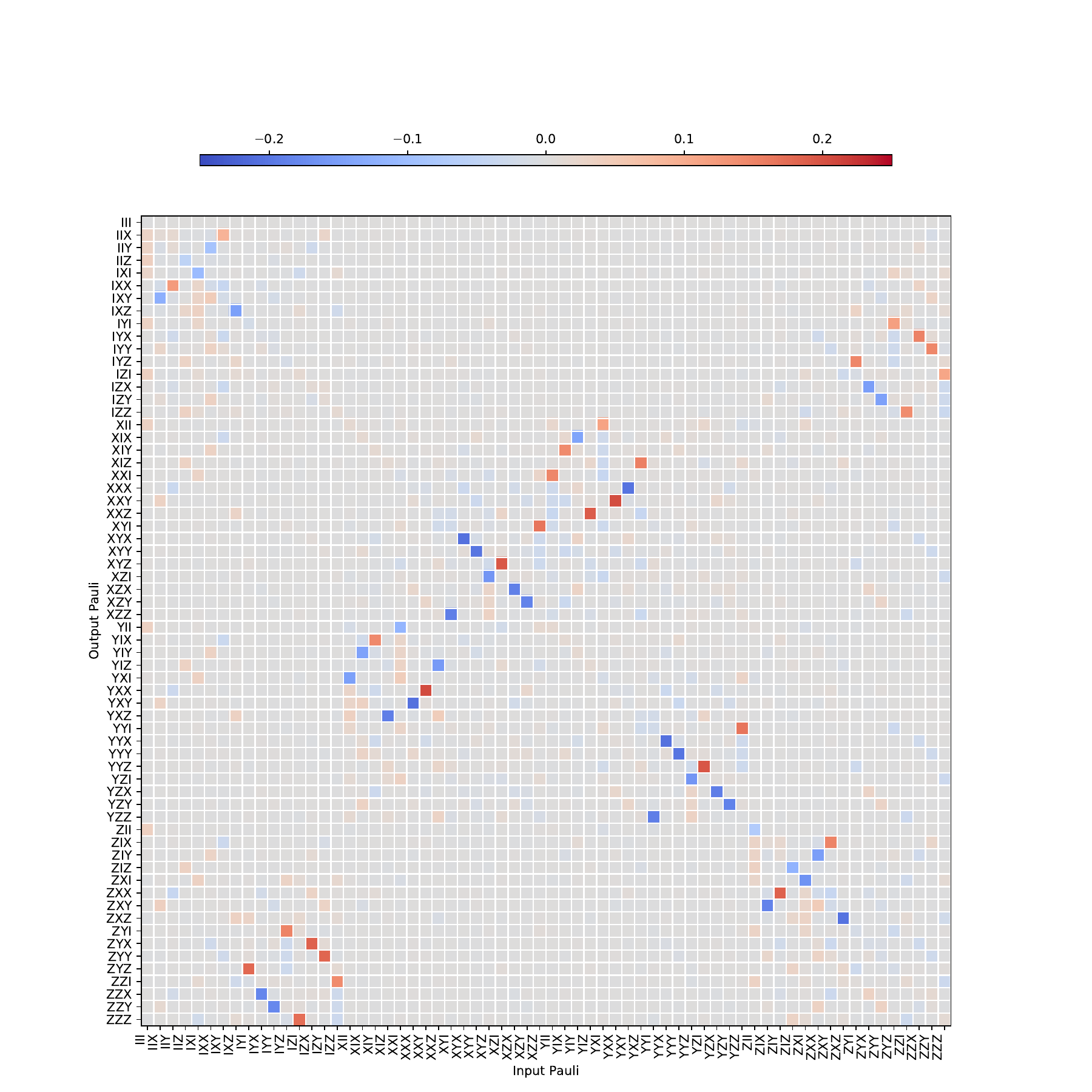}
    \caption{Difference between the Pauli Transfer Matrix of the parallel version of of $R_{ZX}(\pi / 2)$, reconstructed from Process Tomography, and its expected ideal PTM. We have changed the scaling for clarity. In this figure, we have not corrected for SPAM errors; therefore, since the PTM peak values are not exactly $\pm 1$, this figure resembles the previous one, but with inverted colors.}
    \label{fig:pauli_transfer_matrix_diff}
\end{figure*}

Having reconstructed the operator $\Lambda$ for the gate $\Pgate[abc](\pi / 2)$ via maximum-likelihood tomography, we can calculate the PTM, which we present in \cref{fig:pauli_transfer_matrix}. In \cref{fig:pauli_transfer_matrix_diff} we show the difference between the PTM of our implementation of $\Pgate[abc](\pi/2)$ and its ideal PTM. The most significant finding, especially in \cref{fig:pauli_transfer_matrix_diff}, is that the difference to the ideal process reveals a structure similar to the PTM itself. This happens because we have not corrected for SPAM errors, so there is a large difference between the peak values of the PTM and their ideal value of $\pm 1$. Nonetheless, the fidelity reconstructed from this PTM matrix (done indirectly via the Choi matrix) is enough to show the parallelization yields an advantage over serialization, as can be seen in \cref{fig:tomography}.


\end{document}